\documentclass[preprint]{emulateapj}

\usepackage{verbatim}

\def\E{{\cal E}}
\def\F{{\cal F}}
\def\R{{\cal R}}

\newcommand{\simgt}{\,\hbox{\lower0.6ex\hbox{$\sim$}\llap{\raise0.6ex\hbox{$>$}}}\,}
\newcommand{\simlt}{\,\hbox{\lower0.6ex\hbox{$\sim$}\llap{\raise0.6ex\hbox{$<$}}}\,}

\begin{document}

\title{Stirring Up the Pot: Can Cooling Flows In Galaxy Clusters Be Quenched By Gas Sloshing?}

\author{J.A. ZuHone\altaffilmark{1}, M. Markevitch\altaffilmark{1}, R. E. Johnson\altaffilmark{2}$^{,}$\altaffilmark{1}}

\altaffiltext{1}{Smithsonian Astrophysical Observatory, Harvard-Smithsonian Center for Astrophysics, Cambridge, MA 02138}
\altaffiltext{2}{Department of Physics and Astronomy, Wilder Lab, Dartmouth College, Hanover, NH 03755}

\begin{abstract}
X-ray observations of clusters of galaxies reveal the presence of edges in surface brightness and temperature, known as ``cold fronts''. In relaxed clusters with cool cores, these commonly observed edges have been interpreted as evidence for the ``sloshing'' of the core gas in the cluster's gravitational potential. Such sloshing may provide a source of heat to the cluster core by mixing hot gas from the cluster outskirts with the cool core gas. Using high-resolution $N$-body/Eulerian hydrodynamics simulations, we model gas sloshing in galaxy clusters initiated by mergers with subclusters. The simulations include merger scenarios with gas-filled and gasless subclusters. The effect of changing the viscosity of the intracluster medium is also explored, but heat conduction is assumed to be negligible. We find that sloshing can facilitate heat inflow to the cluster core, provided that there is a strong enough disturbance. Additionally, sloshing redistributes the gas in the cluster core, causing the gas to expand and decreasing the efficiency of radiative cooling. In adiabatic simulations, we find that sloshing can raise the entropy floor of the cluster core by nearly an order of magnitude in the strongest cases. If the ICM is viscous, the mixing of gases with different entropies is decreased and consequently the heat flux to the core is diminished. In simulations where radiative cooling is included, we find that though eventually a cooling flow develops, sloshing can prevent the significant buildup of cool gas in the core for times on the order of a Gyr for small disturbances and a few Gyr for large ones. If repeated encounters with merging subclusters sustain the sloshing of the central core gas as is observed, this process can provide a relatively steady source of heat to the core, which can help to prevent a significant cooling flow.
\end{abstract}

\keywords{galaxies: clusters: general --- X-rays: galaxies: clusters --- methods: N-body simulations, hydrodynamic simulations}

\section{Introduction\label{sec:intro}}

Galaxy clusters are filled with hot X-ray emitting gas (ICM), whose
radiative cooling time is much longer than the cluster age (several Gyr)
over most of the cluster volume.  An interesting exception is the cores of
clusters with sharply peaked mass profiles, marked by giant central
elliptical (cD) galaxies (e.g., Jones \& Forman 1984; Peres et al.\ 1998),
which constitute the majority of present-day clusters.  Within $r\sim 100$
kpc of the cluster center, the ICM temperature declines sharply toward the
center (e.g., Fukazawa et al.\ 1994; Kaastra et al.\ 2004), while the gas density increases.  This creates a rather distinct, very X-ray luminous
central region of low-entropy gas with radiative cooling times much shorter
than the cluster age, which makes this region thermally unstable.  This
realization gave rise to a ``cooling flow'' scenario
\citep[e.g.][]{fab77,fab94}, in which the core gas cools, contracts to
maintain its pressure, cools even faster, and eventually turns into stars or
molecular clouds near the center at a rates up to several 100$M_\odot$.
However, the expected large amounts of cold matter were not observed in the
X-ray or other wavelengths, presenting early problems for this model (and
giving rise to complicated solutions such as partial-coverage
self-absorption, e.g., \citet{allenfabian97}). Finally, high-resolution {\it
 XMM-Newton}\/ spectroscopy \citep{pet01,pet03} and {\it Chandra}\/
spectral imaging \citep{dav01} showed that there is indeed little gas below
$T\simeq 1$ keV in the cores of clusters with some of the highest cooling
rates.

Since cooling via X-ray radiation is directly observed, this requires a
compensatory steady heating mechanism.  Proposed candidates include magnetic
field reconnection \citep{sok90}, thermal conduction due to electron
collisions (e.g., Narayan \& Medvedev 2001; Fabian, Voigt, \& Morris 2002;
Zakamska \& Narayan 2003) and turbulent conduction (e.g., Cho et al.\ 2003;
Voigt \& Fabian 2004), and heating by cosmic rays (e.g., Colafrancesco \&
Marchegiani 2008); a recent review can be found in \citet{petersonfabian06}.
The currently favored mechanism is heating by the central AGN (e.g.,
B\"ohringer et al.\ 1993; Binney \& Tabor 1995; McNamara et al.\ 2001, 2005;
Fabian et al.\ 2006; Forman et al.\ 2007; for a recent review see McNamara
\& Nulsen 2006). The AGN explosions blow the ubiquitous bubbles in the ICM
and inject energy into the ICM in the form of relativistic particles as well
as mechanical energy \citep{chu02}.  However, the precise mechanism by which this energy heats the central ICM is still unclear. A fine balance between AGN
explosions and cooling is required to avoid the complete blow-up of the cool
cores, which gave rise to ``feedback'' models, where the cooling flow itself
feeds the AGN.  However, several cooling flow clusters do not contain prominent
bubbles or a presently bright AGN (e.g., A1795, Ophiuchus, A2029).  They may
need other heating mechanisms.

Thermal conduction is a particularly attractive alternative idea, because it
taps the vast reservoir of thermal energy in the gas just outside the cool
core, while automatically ensuring that the core will not be overheated,
since the heat influx decreases with diminishing temperature gradient. The
classic plasma conductivity via Coulomb collisions was shown to be
insufficient even at its full Spitzer value (e.g., Zakamska \& Narayan
2003). It has a strong temperature dependence and decreases right where it
is most needed, and tangled magnetic fields should further suppress it
(as was indeed observed outside the cool cores, Markevitch et al.\ 2003a).

There may be another mechanism to conduct heat from the surrounding hot gas
into the cooling core, which is the subject of this work. {\it Chandra}\/
revealed that the central cool gas in many, if not most, cool-core clusters
is ``sloshing'' in the central potential well, generating the ubiquitous
arc-like gas density discontinuities (``cold fronts''), concentric with
respect to the brightness peak of the cluster (e.g., Mazzotta et al.\ 2001;
Markevitch et al.\ 2001, 2003b; Churazov et al.\ 2003; Ascasibar \&
Markevitch 2006, hereafter AM06; for a review of observations and
simulations see Markevitch \& Vikhlinin 2007).  Such sloshing can be the
result of a recent disturbance of the ICM by, e.g., a subcluster infall or
an AGN explosion.  The kinetic energy of the sloshing gas eventually
dissipates as heat, but probably slowly enough to be insignificant compared
to cooling (Markevitch et al.\ 2001). However, as proposed in Markevitch \&
Vikhlinin (2007), sloshing can also bring the outer, high-entropy gas into
the cool core, facilitating its contact and mixing with the cooling gas and
thus providing a net heat inflow.

\begin{figure*}
\begin{center}
\plotone{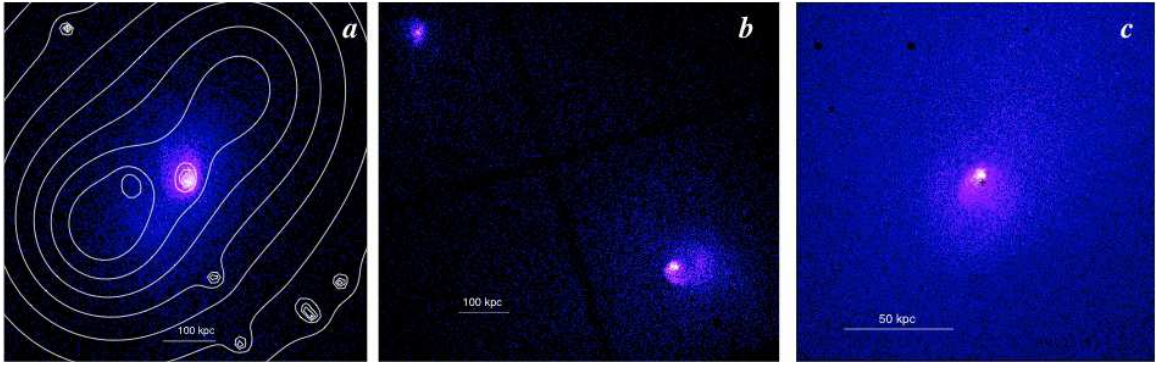}
\caption{Sloshing of the core gas induced by a merger --- sequence of
events (see text). ({\em a}) {\it Chandra} X-ray image of RXJ\,1347--1145
(color) with an overlay of the projected mass map from gravitational lensing
(contours from Miranda et al.\ 2008). The X-ray image shows a sharp cold
front enveloping the core from the south.  ({\em b}) {\it Chandra} image of
A1644; sloshing has started in the core of the southern subcluster.  ({\em c})
{\it Chandra} image of the Ophiuchus cluster, showing several concentric edges
at different radii in the characteristic spiral pattern.  The cross marks
the center of the cD galaxy (and presumably the peak of the gravitational
potential).\label{fig:sb_maps}}
\end{center}
\end{figure*}

\begin{figure*}
\begin{center}
\plotone{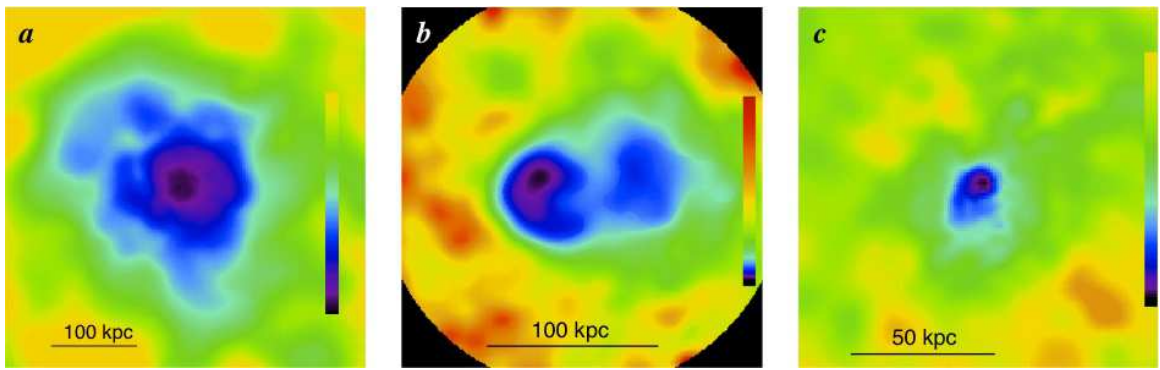}
\caption{{\it Chandra} maps of ``pseudo-entropy'' for clusters with sloshing. ({\em a}) A2204. ({\em b}) The core of the main subcluster of A1644 (see Fig.\ 1{\em b}; reproduced from Johnson et al.\ (2009). ({\em c}) Ophiuchus cluster (see Fig.\ 1{\em c}). The outer, higher entropy gas is making inroads into the cool cores as a result of sloshing, as suggested by the spiral arms and asymmetry of the cores.\label{fig:entr_maps}}
\end{center}
\end{figure*}

When a subcluster passes through the core of a cluster containing the central
cool gas, events unfold along the sequence simulated and discussed in AM06.
We illustrate it here by the real cluster data in Figure \ref{fig:sb_maps}, which shows {\it
 Chandra}\/ X-ray images of clusters RXJ\,1347--1145, A1644 and Ophiuchus.
For RXJ\,1347--1145, a lensing mass overlay (Miranda et al.\ 2008) reveals a
subcluster that apparently has just passed south of the cool core, setting
off sloshing of the core gas, which generated a prominent cold front.  As the disturbing subcluster moves away,
the central gas continues sloshing, developing multiple concentric edges,
often in a spiral pattern, as seen in A1644 (Johnson et al.\ 2010; Lagana et
al.\ 2010) and Ophiuchus (AM06; Million et al.\ 2010). A spiral pattern of
cold fronts may be a natural state for a disturbed rotating stratified
cluster atmosphere \citep{kes10}.  The dense
``spiral arms'' consist of cool gas originating in the core, while the less
dense gas between those arms is the higher-entropy gas brought inside by
sloshing. This is seen in Figure \ref{fig:entr_maps}, which shows maps of
pseudo-entropy%
\footnote{$s\equiv T S_X^{-1/3}$, where $T$\/ is the projected gas
 temperature and $S_X$ is the X-ray brightness. $S_X\sim \rho^2$ in the
 {\it Chandra} energy band. This quantity does not take into account the
 line-of-sight geometry and is only used for qualitative illustrations. The
 maps for Ophiuchus and A2204 were derived in this work from the {\it Chandra}
 data as described in Johnson et al.\ (2009).}
for three clusters with prominent cold fronts in their cores: Ophiuchus,
A2204 (see also Sanders et al.\ 2009), and A1644. Indeed, in the absence of strong
shocks sloshing is a nearly adiabatic process.  If the gas distribution
prior to the disturbance is centrally symmetric with a steep entropy drop
toward the center, the current specific entropy of the gas can be used to
determine its original distance from the center. The maps in Figure \ref{fig:entr_maps} suggest
that some of the higher-entropy gas currently inside the core should have
originated at significantly greater radii.  For Ophiuchus, Million et al.\
(2010) also presented a metallicity map for the core gas, which suggests a
consistent picture (again, in the assumption of an initial centrally
symmetric, peaked metallicity distribution).

Previous investigations have sought to address the question of the ability of mergers to shut or stave off cooling flows \citep[e.g.,][]{fab91,gom02,poo08}, but these works typically assume a context within which the merger destroys the cool core completely through ram-pressure stripping; we are interested in the more subtle and long-lasting effect of the sloshing of the core gas and its mixing with hotter gas from the outskirts, which usually will not destroy the cool core. In this paper, we use hydrodynamical/$N$-body simulations of mergers of
galaxy clusters with small subclusters to determine whether the resulting
sloshing can facilitate an efficient heat inflow to offset runaway cooling
in the cluster core.  In the present work, we do not include collisional
heat conduction; once the high-entropy gas is brought in contact with the
cool core gas, the only mechanism of heat exchange is mixing. In a future
paper, we will include collisional heat conduction as an additional
mechanism.  We employ idealized, binary merger simulations to isolated the physical phenomena of interest. Such simulations have been employed
in many previous works, typically in order either to explore a parameter space of
mergers \citep[e.g.,][]{ric01,poo06,poo07} or to simulate
specific cluster merger scenarios \citep[e.g.,][]{spr07,zuh09}. 

This paper is organized as follows: in \S2, we describe the
characteristics of the simulations and the code. In \S3, we
describe the characteristics of gas sloshing in our simulations and its
effect on the cluster cool core. In \S4, we discuss whether or
not sloshing can be effective in offsetting radiative cooling. Finally, in \S5 we summarize our results and discuss
future developments of this work. We assume a
flat $\Lambda$CDM cosmology with $h = 0.7$, $\Omega_{\rm m} = 0.3$, and
$\Omega_b = 0.02h^{-2}$.

\section{Simulations\label{sec:sims}}

\subsection{Method\label{sec:method}}

Similarly to AM06, we simulate idealized mergers of 2 clusters with mass ratios $R \equiv M_1/M_2$, in the range 5-100, with different impact parameters, and with the infalling subcluster with gas or only dark matter (i.e., a subcluster stripped of its gas during previous interactions). We performed our simulations using FLASH, a parallel hydrodynamics/$N$-body astrophysical simulation code developed at the Center for Astrophysical Thermonuclear Flashes at the University of Chicago \citep{fry00}. FLASH uses adaptive mesh refinement (AMR), a technique that places higher resolution elements of the grid only where they are needed. We are interested in capturing sharp ICM features like shocks and ``cold fronts'' accurately, as well as resolving the inner cores of the cluster dark matter halos. It is particularly important to be able to resolve the grid adequately in these regions. AMR allows us to do so without needing to have the whole grid at the same resolution. 

FLASH solves the Euler equations of hydrodynamics using the Piecewise-Parabolic Method (PPM) of \citet{col84}, which is ideally suited for capturing shocks and contact discontinuties (such as the ``cold fronts'' that appear in our simulations). For simulations including viscosity, it is modeled as a diffusive flux term that is added to the Euler equations. In these simulations, we are more interested in the qualitative effects of including an explicit viscosity, rather than attempting to model the precise nature of the viscosity of the ICM (which we will reserve for future papers). Hence, we have assumed a constant kinematic viscosity equal to the Spitzer value at a radius $\sim$50~kpc. For all our simulations, we assume an ideal equation of state with $\gamma = 5/3$. Though most of our simulations are adiabatic, we have a set of simulations where we have included the effects of radiative cooling. For this, we have used a cooling tables derived from a MeKaL model \citep{MKL95}, assuming a metallicity $Z = 0.8~Z_{\odot}$, a value relevant for the cool cores.

We represent the collisionless dark matter component of galaxy clusters as a set of gravitating particles. For this purpose the FLASH code also includes an $N$-body module which uses the particle-mesh method to map accelerations from the AMR grid to the particle positions. The gravitational potential itself is computed using a multigrid solver included with FLASH \citep{ric08}, in the assumption that both the dark matter and gas components contribute to the mass density for solving the Poisson equation.

\subsection{Initial Conditions\label{sec:ICs}}

Our initial conditions for our clusters in these idealized simulations have been set up in the same manner as in AM06, which we will briefly summarize here. 

For the cluster dark matter profile we have chosen a \citet{her90} profile:
\begin{equation}
\rho_{\rm DM} = \frac{M_0}{2{\pi}a^3}\frac{1}{(r/a)(1+r/a)^3}
\end{equation}
where $M_0$ and $a$ are the scale mass and length of the DM halo. The Hernquist profile shares with the more commonly employed \citet[NFW]{NFW97} profile a ``cuspy'' inner radial dependence of the dark matter density, but results in simpler expressions for the mass, potential, and particle distribution functions. Because we are interested in the consequences of the interaction for only the central regions of the main cluster, the difference in the density dependence for large radii is unimportant. For the gas temperature, we use a phenomenological formula:
\begin{equation}
T(r) = \frac{T_0}{1+r/a}\frac{c+r/a_c}{1+r/a_c}
\end{equation}
where $0 < c < 1$ is a free parameter that characterizes the depth of the temperature drop in the cluster center and $a_c$ is the characteristic radius of that drop. This functional form can reproduce cluster temperature profiles of many observed relaxed galaxy clusters, which have a characteristic temperature drop in the center due to cooling. With this temperature profile, the corresponding gas density can be derived by imposing hydrostatic equilibrium. The baryon fraction is set by the constraint that at large radii it should be constant, which we set to $M_{\rm gas}/M_0 = \Omega_{\rm gas}/\Omega_{\rm DM}$. 

After the radial profiles are determined, it remains to set up the distribution of positions and velocities for the dark matter particles. Here we follow the procedure outlined in \citet{kaz06}. For the particle positions, a random deviate $u$ is uniformly sampled in the range [0,1] and the function $u = M_{\rm DM}(r)/M_{\rm DM}(r_{\rm max})$ is inverted to give the radius of the particle from the center of the halo. For the particle velocities, the procedure is less trivial. Many previous investigations have made use of the ``local Maxwellian approximation.'' In this procedure, at a given radius, the particle velocity is drawn from a Maxwellian distribution with dispersion $\sigma^2(r)$, where the latter quantity has been derived from solving the Jeans equation \citep{bin87}. It has been shown that this approach is not sufficient to accurately represent the velocity distribution functions of dark matter halos with a central cusp such as the NFW profile \citep{kaz04}. To accurately realize particle velocities, we choose to directly calculate the distribution function via the Eddington formula \citep{edd16}:
\begin{equation}
\F(\E) = \frac{1}{\sqrt{8}\pi^2}\left[\int^\E_0{d^2\rho \over d\Psi^2}{d\Psi \over \sqrt{\E - \Psi}} + \frac{1}{\sqrt{\E}}\left({d\rho \over d\Psi}\right)_{\Psi=0} \right]
\end{equation}
where $\Psi = -\Phi$ is the relative potential and $\E = \Psi - \frac{1}{2}v^2$ is the relative energy of the particle. We tabulate the function $\F$ in intervals of $\E$ interpolate to solve for the distribution function at a given energy. Particle speeds are chosen from this distribution function using the acceptance-rejection method. Once particle radii and speeds are determined, positions and velocities are determined by choosing random unit vectors in $\Re^3$. All of our simulations employ no fewer than ${\sim}2 \times 10^7$ particles for representing the dark matter.

Our merging clusters consist of a large, ``main'' cluster, and a small infalling subcluster. They are characterized by the mass ratio $R \equiv M_1/M_2$, where $M_1 = M_0R/(1+R)$ and $M_2 = M_0/(1+R)$ are the masses of the main cluster and the infalling satellite, respectively. The total cluster mass $M_0$ for each simulation is set to $1.5 \times 10^{15} M_\odot$. To scale the initial profiles for the various mass ratios of the clusters, the combinations $M_i/a_i^3$, $c_i$, and $a_{c,i}/a_i$ are held constant. For the main cluster, we chose $a_1$ = 600~kpc, $c_1$ = 0.17, and $a_{c,1}$ = 60~kpc, to resemble mass, gas density, and temperature profiles typically observed in real galaxy clusters. In particular, our main cluster closely resembles A2029 \citep[e.g.,][]{vik05}, a hot, relatively relaxed cluster with sloshing in the cool core.

\begin{figure}
\begin{center}
\plotone{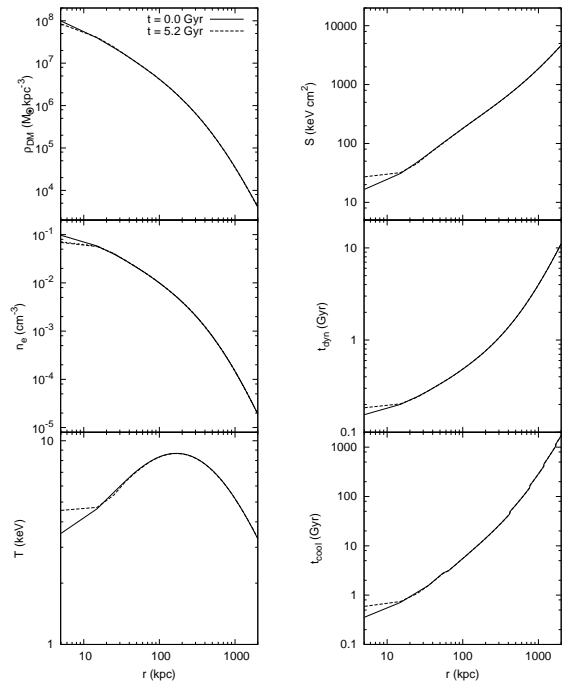}
\caption{Radial profiles of dark matter density, electron number density, gas temperature, entropy, dynamical time, and cooling time of the single-cluster test simulation at $t$ = 0.0~Gyr and at the epoch $t$ = 5.2~Gyr.\label{fig:stable}}
\end{center}
\end{figure}

For all of the simulations, we set up the two clusters within a cubical computational   domain of width $L = 10$ Mpc on a side. Both objects start at a separation of $d$ = 3~Mpc, and with an initial impact parameter $b$ that we may vary for the differing simulations. The initial cluster velocities are chosen so that the total kinetic energy of the system is set to a fraction 0 $\leq K \leq$ 1 of its potential energy, approximating the objects as point masses:
\begin{equation}
E \approx (K-1)\frac{GM_1M_2}{d} = (K-1)\frac{R}{(1+R)^2}\frac{GM_0^2}{d}
\end{equation} 
So the initial velocities in the reference frame of the center of mass are set to
\begin{equation}
v_1 = \frac{R\sqrt{2K}}{1+R}\sqrt{\frac{GM_0}{d}} ; v_2 = \frac{\sqrt{2K}}{1+R}\sqrt{\frac{GM_0}{d}}
\end{equation}
For all of our simulations, we have set $K = 1/2$. 

To test the robustness of our initial model for the clusters we perform a single-cluster test. Figure \ref{fig:stable} shows the profiles of dark matter density, gas density, gas temperature, and gas entropy (defined here as $S \equiv k_BTn_e^{-2/3}$) at the beginning of the simulation and at a later epoch, demonstrating the stability of the cluster at all radii excepting the innermost couple of zones (of width $\sim$~5~kpc) due to force smoothing (a known numerical effect due to the inability to resolve the gravitational force on scales smaller than the grid resolution). In particular, it is important to note that although the entropy in these zones varies by about $\sim 50\%$ from the initial value, this deviation is insignificant when compared to the entropy generated by the heating of the cluster core by subcluster passages, as will be shown below. 

\subsection{Likelihood of Merger Scenarios\label{sec:likelihood}}

\begin{table}[thdp]
\tabletypesize{\scriptsize}
\caption{Merger Statistics During the Last 6~Gyr\label{tab:merger_stats}}
\begin{center}
\begin{tabular}{ccc}
\hline
 & Number of Mergers with $\R < R$ & \\
\hline
\hline
$R$ & & $N(\R < R)$ \\
\hline
5     & & 1 \\
20   & & 3 \\
100 & & 8 \\
\hline
 & Probability of Merger with $V < v_t$ & \\
\hline
\hline
$b$ (kpc) & $v_t/V_C$ & $P(V < v_t)$ \\
\hline
50     & 0.018 & 0.017 \\
200   & 0.072 & 0.070 \\
500   & 0.186 & 0.183 \\
1000 & 0.358 & 0.333 \\
\hline
\end{tabular}
\end{center}
\end{table}

\begin{table*}[thdp]
\caption{Initial Merger Parameters\label{tab:SimGrid}}
\begin{center}
\begin{tabular}{cccccccc}
\hline
\hline
Simulation & $R$ & $b$ (kpc) & $v_t/V_C$ & $\nu$ (cm$^2$ s$^{-1}$) & Subcluster Gas? &
Cooling? & $t_0$\\
\hline
R5b500         & 5     & 500   & 0.126 & 0.0 & NO & NO & N/A \\
R5b500v       & 5     & 500   & 0.126 & $1.266 \times 10^{29}$ & NO & NO & N/A \\
R20b200       & 20   & 200   & 0.065 & 0.0 & NO & NO & N/A \\
R20b200v     & 20   & 200   & 0.065 & $1.266 \times 10^{29}$ & NO & NO& N/A \\
R20b200g     & 20   & 200   & 0.065 & 0.0 & YES & NO & N/A \\
R20b200gv   & 20   & 200   & 0.065 & $1.266 \times 10^{29}$ & YES & NO& N/A \\
R100b50       & 100 & 50     & 0.017 & 0.0 & NO & NO & N/A \\
R20b1000g   & 20   & 1000 & 0.325 & 0.0 & YES & NO & N/A \\
R5b500c       & 5     & 500   & 0.126 & 0.0 & NO & YES & 1.733\\
R5b500vc     & 5     & 500   & 0.126 & $1.266 \times 10^{29}$ & NO & YES & 1.733 \\
R20b200gc   & 20   & 200   & 0.065 & 0.0 & YES & YES & 1.445 \\
R20b1000gc & 20   & 1000 & 0.325 & 0.0 & YES & YES & 1.622 \\
\hline
\end{tabular}
\end{center}
\end{table*}

An important question to address regarding our idealized merger simulations is the likelihood of our chosen merger conditions in the real cluster population. We have explored several combinations of mass ratio and impact parameter that would be expected to induce sloshing within the cluster core. This means that for higher mass ratios, we have chosen smaller impact parameters, to ensure that the resulting interaction is strong enough. To determine the likelihood of such merger configurations, we consulted studies of cosmological simulations that determined the statistical properties of galaxy cluster mergers. \citet{fak08} constructed merger trees from the Millenium Simulation \citep{spr05} to quantify the merger rate over a range of descendant halo mass, progenitor mass ratio, and redshift. They found a universal fitting formula for the mean merger rate per halo that is accurate to 10-20\%. From their results, we can make reasonable estimates of the likelihood of finding a merger with a subcluster with a mass ratio $\leq R$ out of all the mergers that occur. Similar investigations of the statistical properties of tangential velocities of subclusters with respect to their merging progenitors have also been carried out. \citet{vit02} and \citet{ben05} examined the tangential and radial components of subcluster velocities over a range of mass ratios. The former found that the distribution of radial and tangential velocities of subclusters for high mass ratios ($R \geq 3$, the range of mass ratios that includes all of the simulations presented in this paper) is well-described by a multivariate Gaussian distribution with the anisotropy parameter $\beta = 1 - \frac{\sigma_t^2}{2\sigma_r^2} \approx 0.6$. From this distribution, we can determine the likelihood of finding a merger with a tangential velocity smaller than (or, equivalently, an impact parameter smaller than) the chosen value. 

Table \ref{tab:merger_stats} shows these two sets of statistics, which are independent of one another. The top of the table shows the expected number of encounters of our single cluster of our chosen initial mass of $M_0 = 1.5 \times 10^{15} M_{\odot}$ with subclusters with a mass ratio $R$ greater than the given value during the past 6~Gyr, computed from integrating Equation 12 from \citet{fak08}. On the bottom part of the table, we show the selected impact parameters (and their corresponding tangential speeds, scaled to the circular velocity at the virial radius $V_c = GM_{\rm vir}/r_{\rm vir}$ as in \citet{vit02}) from our simulations. With these we list the corresponding probability of a single cluster to have an encounter with a tangential velocity less than or equal to the chosen speed, using the distribution noted above given in \citet{vit02}, which for mass ratios $R > 3$ such as our set of simulations is independent of the mass ratio itself.  These two independent sets of statistics indicate we have chosen initial mass ratios and impact parameters that are realistic. Additionally, it is known from observations that sloshing occurs in most cool core clusters \citep{MVF03}. If sloshing is caused by mergers such as those we simulated, which appears to be supported by observations (e.g. Johnson et al.~2009), this in an indirect indication that such mergers are likely to be frequent. 

Table \ref{tab:SimGrid} presents the details of each simulation, including the initial orbital parameters, the value of the viscosity, whether or not the subcluster included gas, and whether or not the effects of cooling are included. 

\section{Results\label{sec:results}}

\subsection{Description of Sloshing\label{sec:descript}}

First, we will briefly describe the sloshing process due to subcluster mergers as elucidated in Section 3 of AM06. Since the details of the process are qualitatively different depending on whether or not the merging subcluster contains gas, we will consider these cases separately.

\begin{figure*}
\begin{center}
\plotone{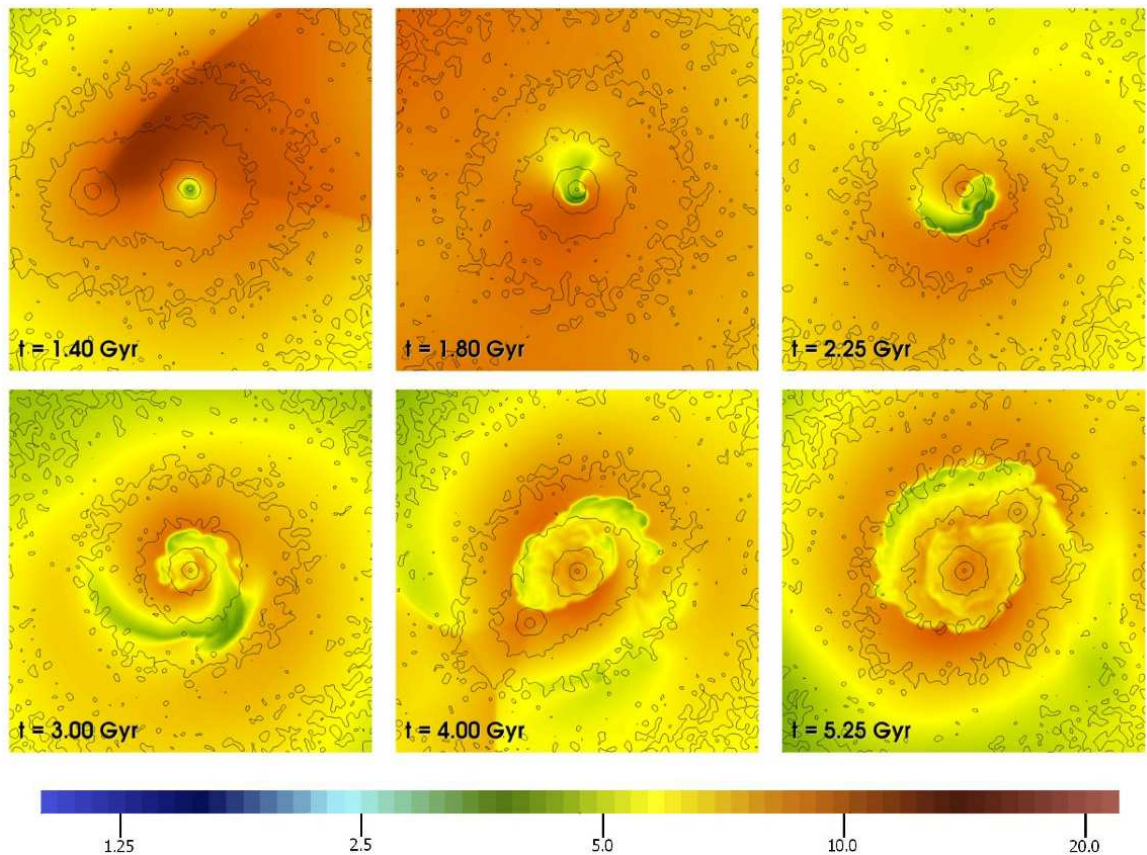}
\caption{Slices through the gas temperature in keV for simulation R5b500 with dark matter contours overlaid. Each panel is 1 Mpc on a side.\label{fig:R5_b500}}
\end{center}
\end{figure*}

\begin{figure*}
\begin{center}
\plotone{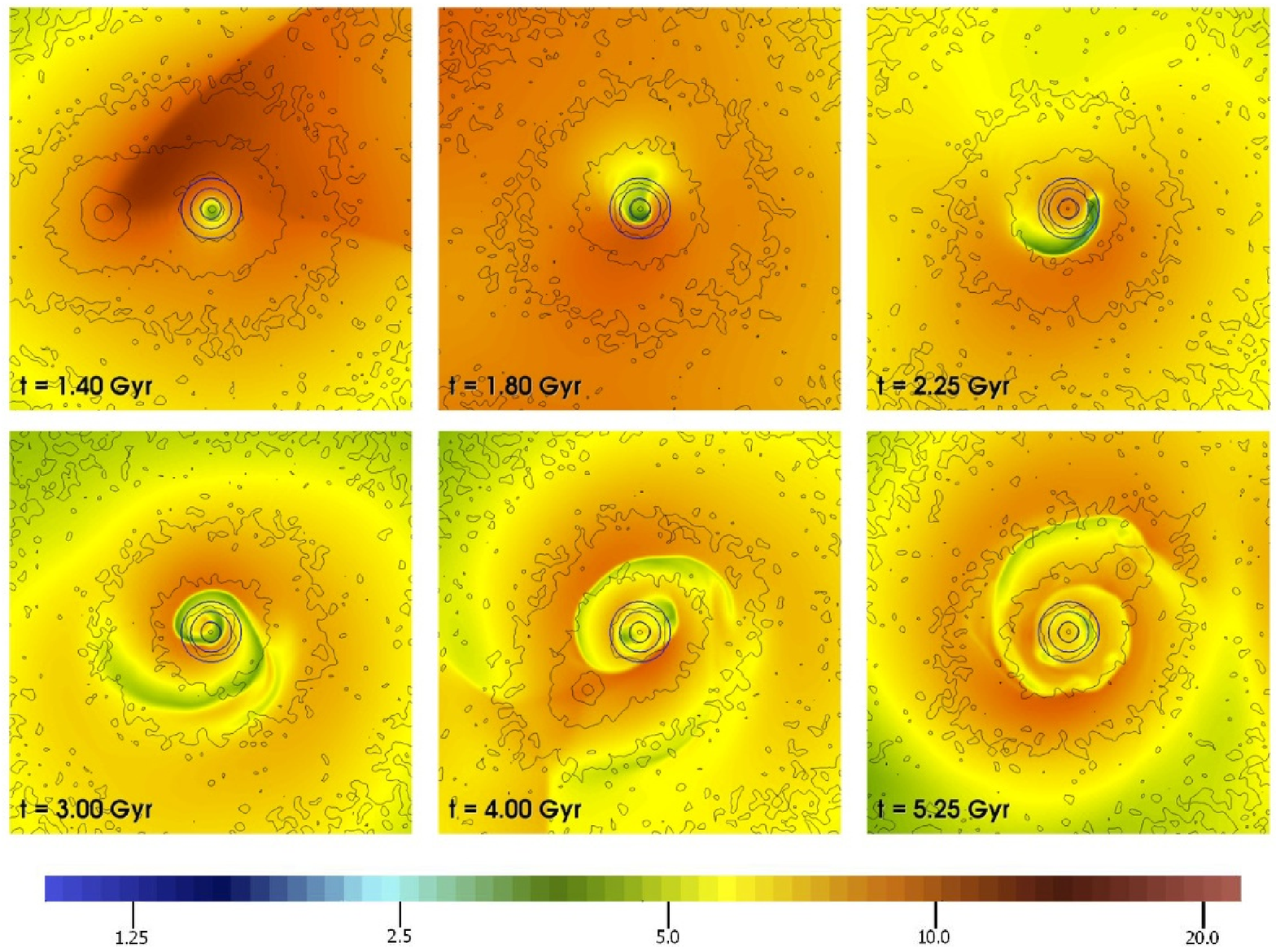}
\caption{Slices through the gas temperature in keV for simulation R5b500v with dark matter contours overlaid. The blue circles mark the radii of 25, 50, and 75~kpc from the gravitational potential minimum. Each panel is 1 Mpc on a side.\label{fig:R5_b500_visc}}
\end{center}
\end{figure*}

In the gasless subcluster cases (simulations R5b500, R5b500v, R20b200, R20b200v, and R100b50), it is assumed that the subcluster has lost its gas due to ram pressure stripping from an earlier phase of the merger (although as the subcluster approaches the main cluster it begins to drag some of the cluster's ICM in a trailing sonic wake, see the first panel of Figures \ref{fig:R5_b500} and \ref{fig:R5_b500_visc}).  The first core passage occurs at approximately $t \sim 1.3$~Gyr after the beginning of the simulation; each simulation is followed until $t = 6.0$~Gyr. As the subcluster approaches the main cluster's core and makes its first passage, the gas and DM peaks of the main cluster feel the same gravity force toward the subcluster and move together towards it. However, the gas feels the effect of the ram pressure of the ambient medium. This fact becomes significant as the gas core is held back from the core of the dark matter by this pressure.After the passage of the core, when the direction of the gravitational force quickly changes, there is a rapid decline of the ram pressure. As a result from this change, the gas core experiences a ``ram pressure slingshot'' \citep{hal04}, where the gas that was previously held back by the ram pressure falls into the DM potential minimum and overshoots it. In addition to the gravitational disturbance, the wake trailing the subcluster transfers some of the angular momentum from the subcluster to the core gas and also acts to help push the core gas out of the DM potential well.

As the cool gas from the core climbs out of the potential minimum, it expands adiabatically. However, the lowest entropy gas quickly begins to sink back towards the potential minimum against the ram pressure from the surrounding ICM. Once again, as the cool gas falls into the potential well it overshoots it, and the process repeats itself.  Each time, a contact discontinuity (``cold front'') is produced. Due to the angular momentum transferred from the subcluster by the wake these fronts have a spiral-shaped structure. Throughout this process, higher-entropy gas from larger radii is brought into contact with the lower entropy gas from the core, and as these gases mix, the entropy of the core gas is increased. 

\begin{figure*}
\begin{center}
\plotone{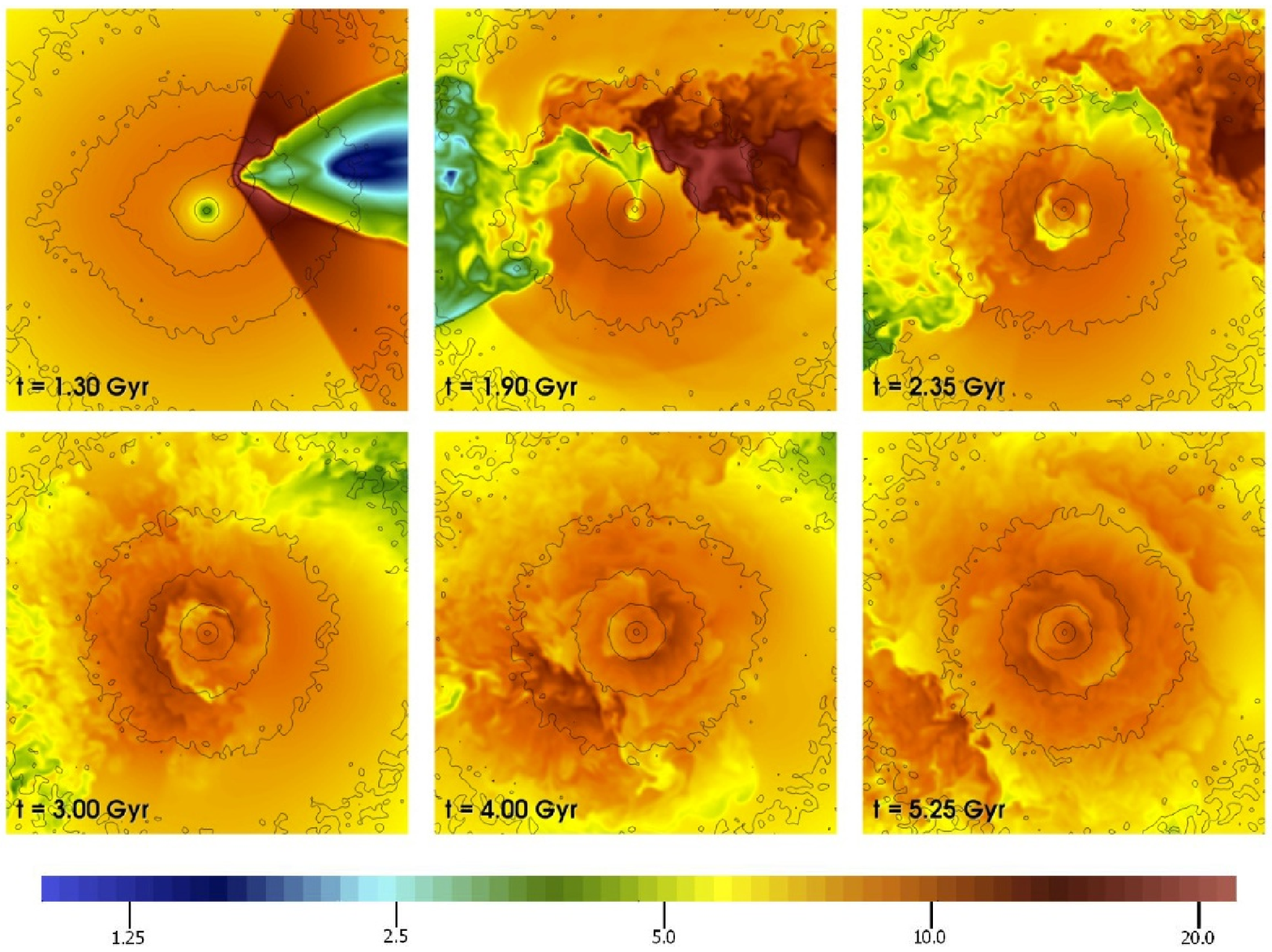}
\caption{Slices through the gas temperature in keV for simulation R20b200g with dark matter contours overlaid. Each panel is 1 Mpc on a side.\label{fig:R20_b200_gas}}
\end{center}
\end{figure*}

\begin{figure*}
\begin{center}
\plotone{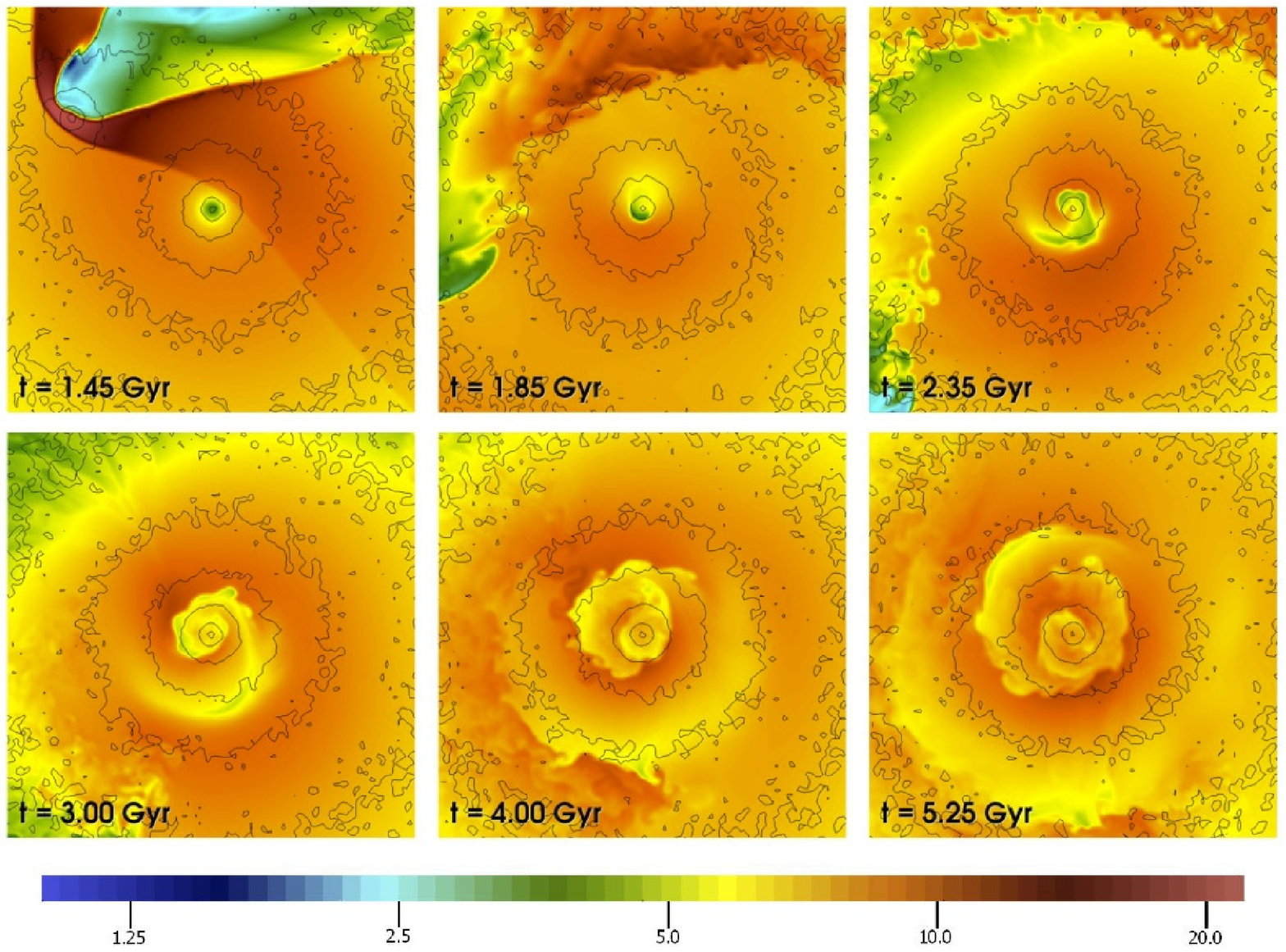}
\caption{Slices through the gas temperature in keV for simulation R20b1000g with dark matter contours overlaid. Each panel is 1 Mpc on a side.\label{fig:R20_b1000_gas}}
\end{center}
\end{figure*}

In the case of a subcluster with gas (simulations R20b200g, R20b200gv, and R20b1000g), instead of a sonic wake, a shock front forms in front of the subcluster as it approaches the main cluster (see Figures \ref{fig:R20_b200_gas} and \ref{fig:R20_b1000_gas}). This has two effects on the core of the main cluster: first, there is an increase in entropy due to the shock as it passes the cluster core, and the shock itself adds a source of pressure to push the cool gas out of the cluster core at an earlier stage compared to the corresponding gasless subcluster case. In cases where the subcluster makes a sufficiently close passage to the cluster core (simulations R20b200g and R20b200gv), the cool core of the main cluster is disrupted completely as it is displaced by the gas from the subcluster (see the second panel of Figure \ref{fig:R20_b200_gas}). This gas mixes in with the core gas of the main cluster and additionally increases the entropy of the core. 

Though the sequence of events and the sloshing pattern itself in our simulations are generally the same as in the simulations of AM06, there are some significant differences, that can be readily seen if our Figures \ref{fig:R5_b500} through \ref{fig:R20_b1000_gas} are compared to the temperature slice figures of that paper. In AM06, the sloshing cold fronts in all of the simulations are all smooth and well-defined spirals, and the cool cores are generally intact after the encounter with the subcluster. By contrast, in our simulations, the cold front surfaces are disrupted by fluid instabilities and the initially cool, dense cores have been replaced by warm, low-density cores, even though we have used the same physical setup and the parameters for some of our simulations are identical to the ones used in AM06 (e.g., $R$ = 5, $b$ = 500 kpc). These crucial differences have to do with the different ways that Gadget-2, a Lagrangian SPH code, versus FLASH 3, an Eulerian AMR code, implement the equations of hydrodynamics. We will elaborate on this difference and its implications in Section \ref{sec:viscosity}.

\subsection{Sloshing in Adiabatic Mergers\label{sec:feedback}}

\begin{figure*}
\begin{center}
\plotone{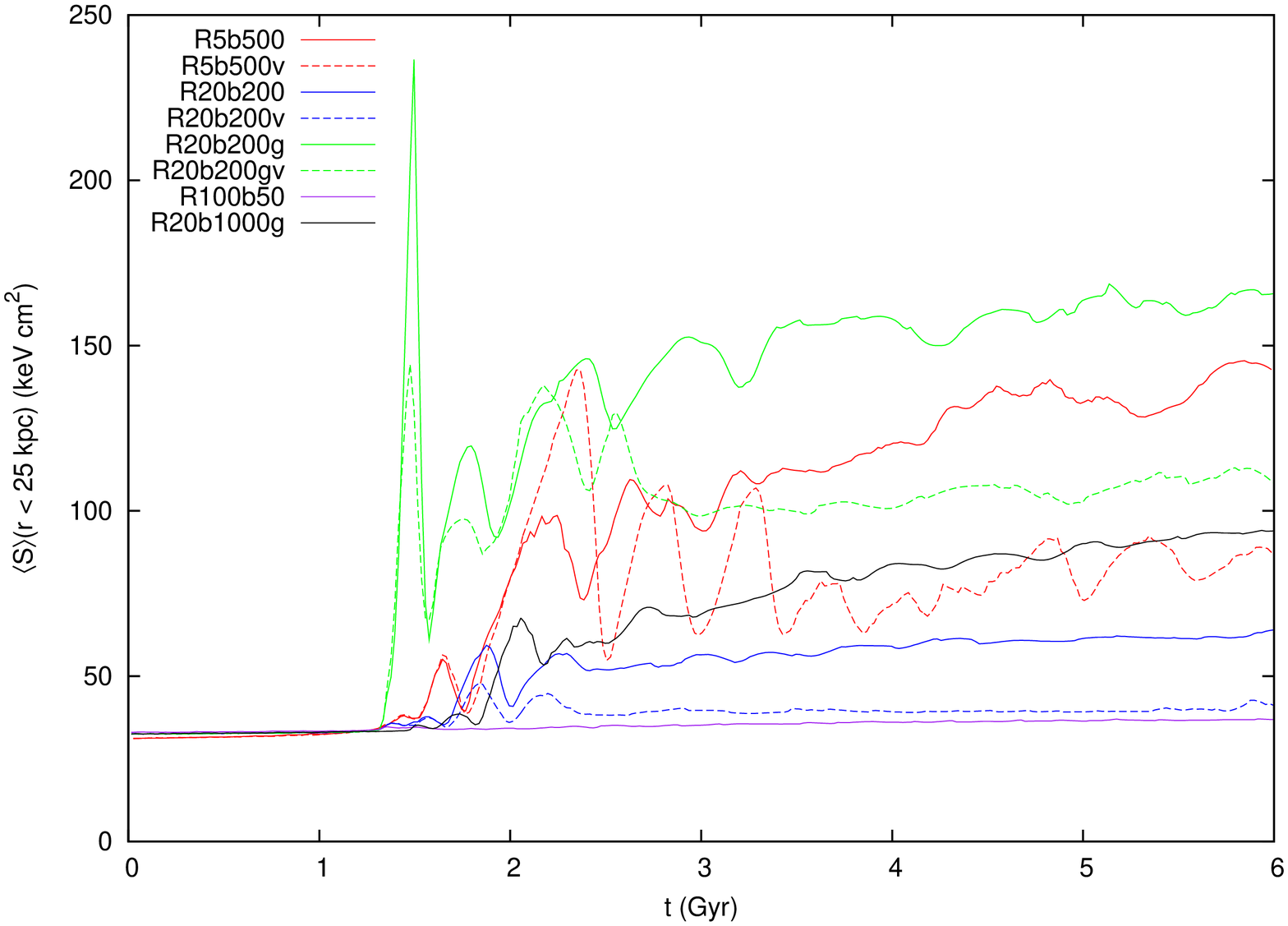}
\caption{Average entropy per unit mass within $r < 25$ kpc for the adiabatic simulations vs. time.\label{fig:entr_vs_time}}
\end{center}
\end{figure*}

To have a clear separation of the various physical effects, and to get a good measure of the amount of heating that may be expected from sloshing, we first investigate the heat generated in merger scenarios that are adiabatic (where radiative cooling is not included in the simulations). As previously mentioned, the first and most important effect is that the specific entropy of the core should increase. Figure \ref{fig:entr_vs_time} shows the evolution of the average specific entropy within a radius of $r < 25$~kpc (the typical radius within which radiative cooling is strong) of the potential minimum of the main cluster. In each simulation, there is an initial transient increase of the entropy per unit mass due to the passage of the wake or shock, but following this there is a more gradual increase due to sloshing. The physical reason for this increase is the inward flow of high-entropy gas from the outer regions, which then mixes with the cool gas. By the end of each simulation ($t$ = 6~Gyr) this increase gradually levels off as the sloshing subsides. The entropy increase is generally stronger in simulations where gas is present in the subcluster, the gravitational interaction with the subcluster is strong (i.e., when the subcluster is more massive and/or passes close to the cool core), and when the viscosity of the ICM is low, due to its effect of suppressing mixing. However, in every case, there is at least a small increase in the core entropy. In this figure it can be seen that for some simulations (particularly R5b500v) the core entropy within $r < 25$~kpc oscillates wildly; this is due to coherent clumps of low-entropy gas sloshing in and out of the volume. The blue circles on Figure \ref{fig:R5_b500_visc} mark the radii of 25, 50, and 75~kpc to show more explicitly how this occurs.

\begin{figure}
\begin{center}
\plotone{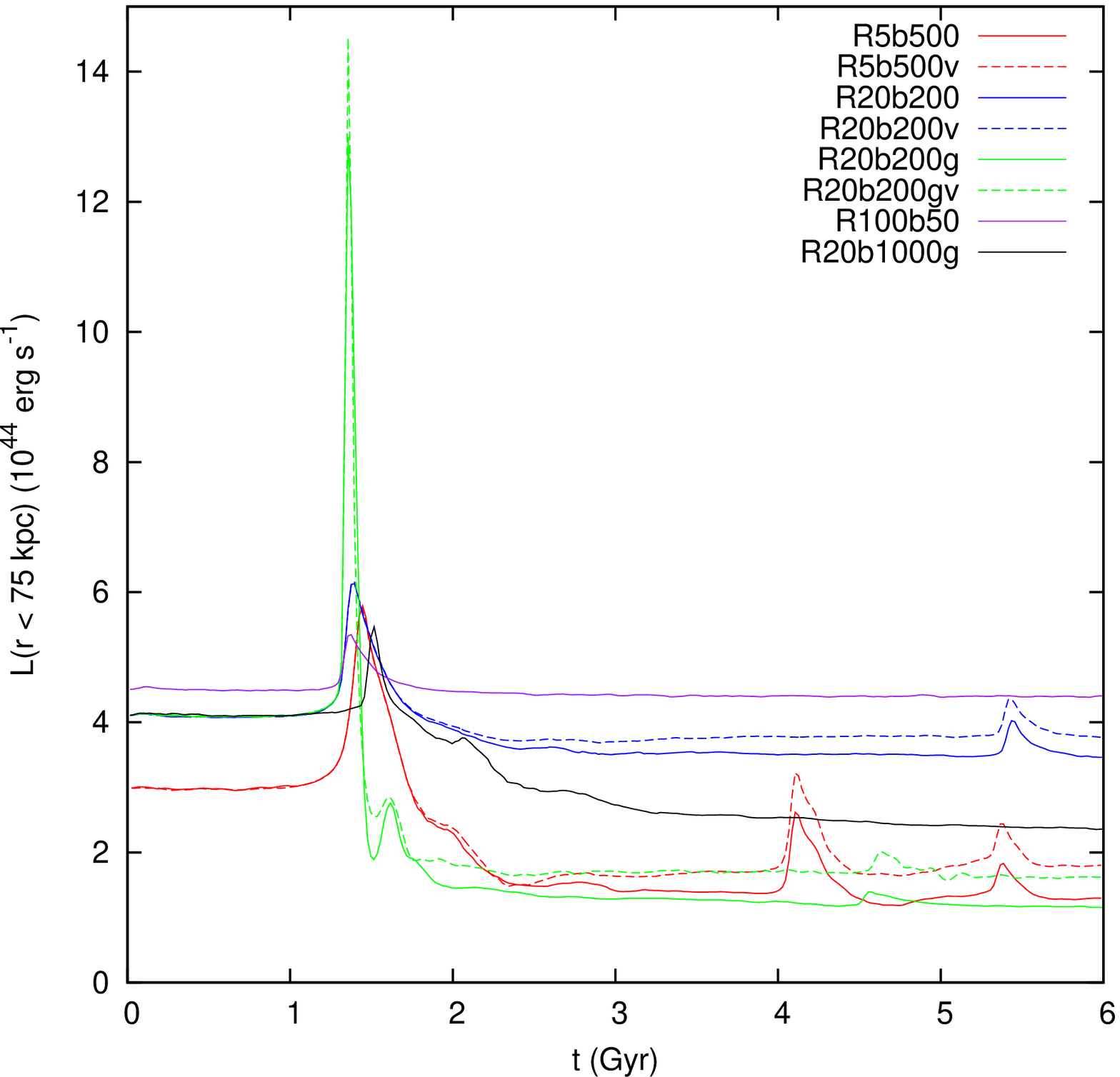}
\caption{Luminosity within $r < 75$ kpc for the adiabatic simulations vs. time.\label{fig:L_vs_time}}
\end{center}
\end{figure}

\begin{figure}
\begin{center}
\plotone{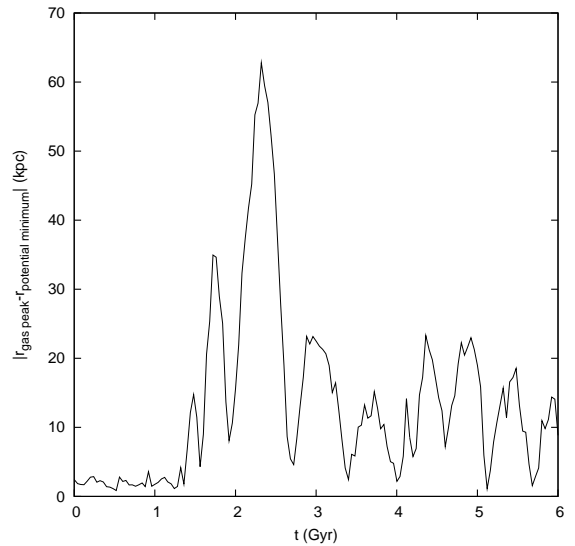}
\caption{Distance between the galaxy cluster potential minimum, and the ``gas peak'' (defined as the center of mass of gas with $t_{\rm cool} \leq 3~{\rm Gyr}$) vs. time, for simulation R5b500v.\label{fig:diff_peaks}}
\end{center}
\end{figure}

\begin{figure}
\begin{center}
\plotone{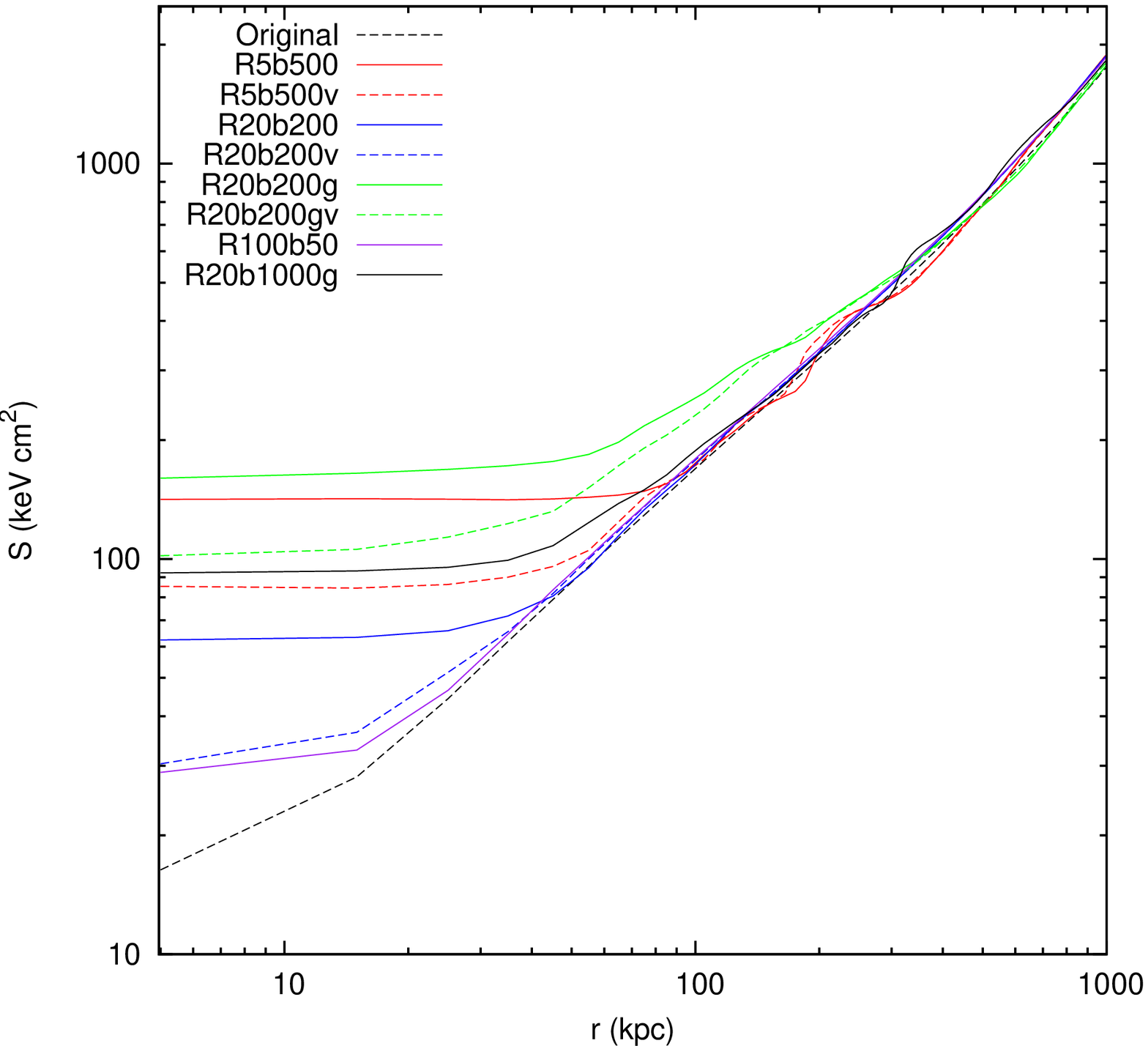}
\caption{Final entropy profiles (at $t$ = 6.0 Gyr) for the adiabatic simulations, compared to the initial profile.\label{fig:e_profiles}}
\end{center}
\end{figure}

A secondary result of sloshing is that as the core entropy is raised, the gas is redistributed in such a way as to decrease cooling. Specifically, the gas expands and the temperature is raised. Since the dependence of the emission on temperature is generally weak (approximately $L_X \propto T^{1/2}$ for T $\simgt$ 3~keV, with a reverse dependence for T $\simlt$ 1~keV), and the dependence on density is strong ($L_X \propto \rho^2$), the net effect is always to decrease cooling. Figure \ref{fig:L_vs_time} shows the evolution of the luminosity within a radius of $r < 75$~kpc (which encompasses most cool cores) for each of the adiabatic simulations. The luminosity within a given radius is constant before the interaction with the subcluster, and after a brief period of increase due to the gas compression from the increased gravitational potential when the subcluster passes by, the luminosity decreases as the gas is being redistributed, a process which takes $\sim$1-2~Gyr. After this time, the luminosity is relatively constant unless there is a second passage of the subcluster, in which case there is a smaller transient increase of luminosity (because the subcluster's dark matter has undergone significant tidal stripping), after which the luminosity settles back to the previous value. Simulations with a greater degree of sloshing result in a greater decrease in luminosity. We show the resulting percentage change in luminosity before and after the encounter with the subcluster within a few radii for the adiabatic simulations in Table \ref{tab:deltaL}.

\begin{table}[thdp]
\caption{Reduction in Cooling: $L_{X,{\rm final}}/L_{X,{\rm initial}}$\label{tab:deltaL}}
\begin{center}
\begin{tabular}{cccc}
\hline
\hline
Simulation & $r < 25~{\rm kpc}$ & $r < 50~{\rm kpc}$ & $r < 75~{\rm kpc}$ \\
\hline
R5b500       & 17\% & 33\% & 48\% \\
R5b500v     & 29\% & 49\% & 62\% \\
R20b200     & 48\% & 76\% & 86\% \\
R20b200v   & 78\% & 89\% & 93\% \\
R20b200g   & 11\% & 21\% & 30\% \\
R20b200gv & 18\% & 32\% & 41\% \\
R100b50     & 89\% & 94\% & 98\% \\
R20b1000g & 28\% & 48\% & 60\% \\
\hline
\end{tabular}
\end{center}
\end{table}

\begin{table}[thdp]
\caption{Ratio of Heating to Cooling, Averaged Over $\sim$5~Gyr of Sloshing $\langle\dot{Q}\rangle/{\langle}L_X{\rangle}$\label{tab:heat_to_cool}}
\begin{center}
\begin{tabular}{cccc}
\hline
\hline
Simulation & $r < 25~{\rm kpc}$ & $r < 50~{\rm kpc}$ & $r < 75~{\rm kpc}$ \\
\hline
R5b500       & 76\% & 60\% & 45\% \\
R5b500v     & 47\% & 34\% & 25\% \\
R20b200     & 19\% & 8\% & 4\% \\
R20b200v   & 6\% & 3\% & 2\% \\
R20b200g   & 314\% & 211\% & 141\% \\
R20b200gv & 173\% & 93\% & 51\% \\
R100b50     & 1\% & 0.2\% & 0.1\% \\
R20b1000g & 41\% & 28\% & 25\% \\
\hline
\end{tabular}
\end{center}
\end{table}

For the purposes of understanding the effect of the displacement of cold gas on the behavior of entropy and luminosity within a given radius of the cluster potential minimum, it is instructive to see the behavior of this displacement with time. Figure \ref{fig:diff_peaks} shows the evolution of the displacement between the gravitational potential minimum of the main cluster and the gas peak of simulation R5b500v, chosen as an example since it represents the scenario with the largest displacement of gas from the center out of all our ``pure sloshing'' simulations (where the gas core is not destroyed as in simulations R20b200g and R20b200gv). The gas peak has been defined as the mass of gas with a cooling time $t_{\rm cool} \leq 3~{\rm Gyr}$, which for our initial cluster roughly corresponds to the gas within our ``cooling radius'' $a_c$. In this case, the first few swings of the sloshing motion bring the low-entropy gas far from the potential minimum, out to nearly $\sim 65$~kpc at maximum displacement (at $t \sim 2.3$~Gyr), but the subsequent maximum displacements of the sloshing motions decrease, falling roughly within $\sim 25$~kpc at all times following the beginning of the sloshing motions. This implies that for a brief interval of time at the onset of sloshing, the evolution of entropy and luminosity within a given radius is dominated by the displacement of dense, low-entropy gas from the center, but this interval is short-lived and subsequent evolution is indicative of the mixing and redistribution of gas within the cluster core.

The end result of the sloshing process on the core can be seen more explicitly by comparing the final entropy profiles ($\sim$5~Gyr after the first core passage) to the initial profile of the main cluster. Figure \ref{fig:e_profiles} shows the final radial entropy profiles for our eight adiabatic simulations, compared with the initial profile of the main cluster. In each case, the entropy profile of the core has been raised and flattened considerably, either by a factor of $\sim$2 in the weaker mergers (R20b200v and R100b50) to a factor of $\sim$6-10 in the stronger cases (R5b500, R5b500v, R20b200, R20b200g, R20b200gv, and R20b1000g), a complete disruption of the cooling flow. These simulations demonstrate that sloshing driven by mergers is very capable of raising the central entropy of a cluster with an initial cool-core configuration to significantly higher levels, at least in the absence of the effects of radiative cooling on the behavior of the gas. 

A comparison of the heating rate from sloshing with the cooling rate is necessary to determine whether or not sloshing is effective in combating the effects of cooling in the core. We measure the instantaneous heating rate within a spherical volume $V$ centered on the main cluster's potential minimum by computing the quantity 
\begin{equation}
\dot{Q} = \int_V{{\rho}T\frac{{\partial}s}{{\partial}t}}dV
\end{equation}
where $\dot{Q}$ is the heating rate in erg s$^{-1}$ and $s = c_{\rm V}\ln{(P\rho^{-\gamma})}$ is the entropy per unit mass of the gas in erg K$^{-1}$ g$^{-1}$. 
For the instantaneous cooling rate within a spherical volume, we integrate over the cooling function:
\begin{equation}
L_X = \int_V{n_en_p\Lambda(T,Z)}dV
\end{equation}
where $L_X$ is the cooling rate in erg s$^{-1}$, $n_e$ is the electron number density in cm$^{-3}$, $n_p$ is the proton number density in cm$^{-3}$, and $\Lambda(T,Z)$ is the cooling function in erg cm$^3$ s$^{-1}$ (where again the cooling function is interpolated from the MeKaL table with an abundance of $Z = 0.8 Z_{\odot}$. These two quantities are computed as a function of time and then averaged over the interval of sloshing. We show the ratio of heating to cooling for each adiabatic simulation within certain specific volumes in Table \ref{tab:heat_to_cool}. In this table it can be seen that the average heat input for all simulations except simulations R20b200g and R20b200gv is less than the average cooling rate over the same interval of time, though for some parameter combinations (e.g., simulations R5b500, R5b500v, and R20b1000g) the average heating rate is a significant fraction of the cooling rate.

\subsection{Sloshing in Mergers With Cooling\label{sec:withcooling}}

\begin{figure*}
\begin{center}
\plotone{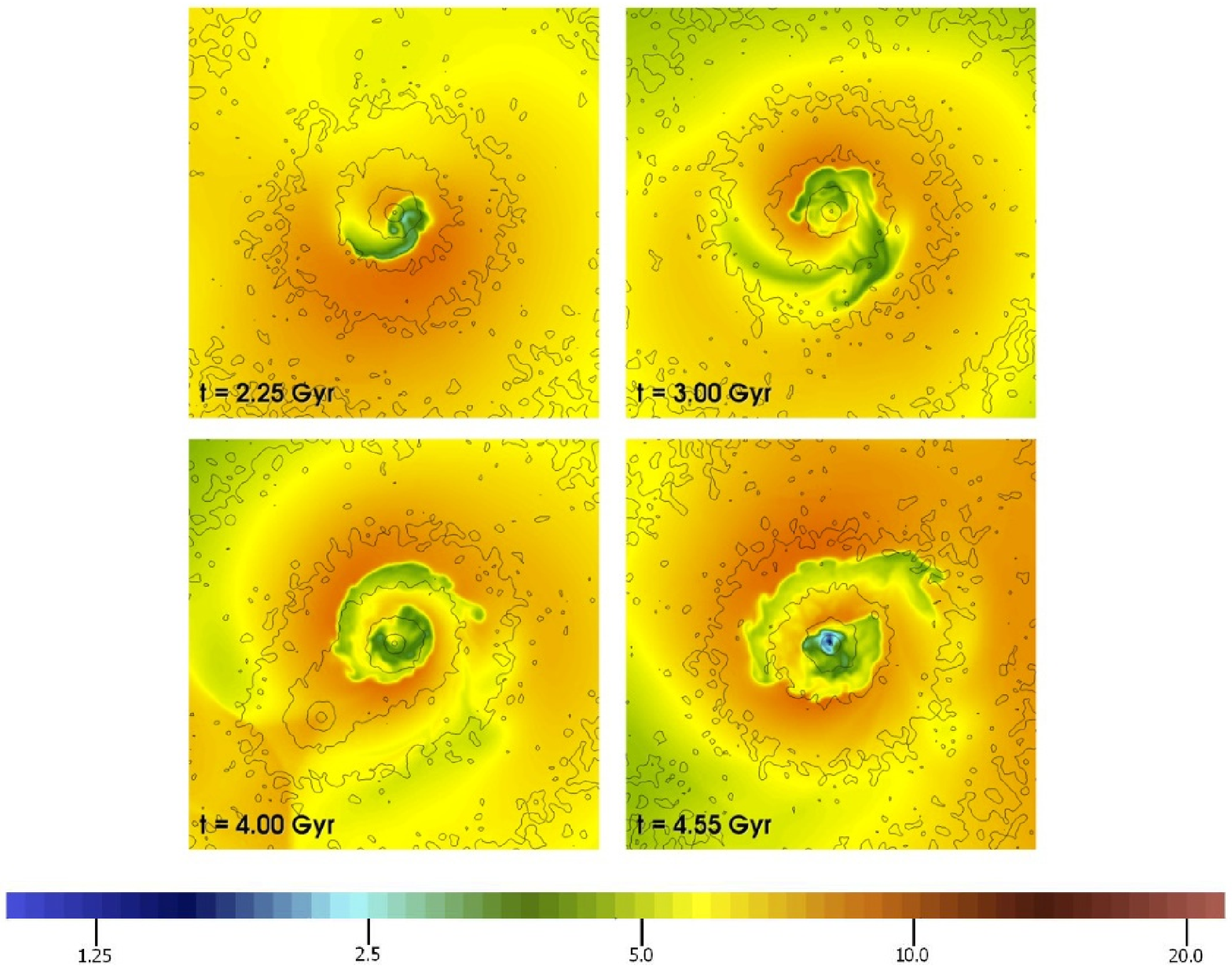}
\caption{Slices through the gas temperature in keV for simulation R5b500c with dark matter contours overlaid. Each panel is 1 Mpc on a side.\label{fig:R5_b500_cool}}
\end{center}
\end{figure*}

\begin{figure*}
\begin{center}
\plotone{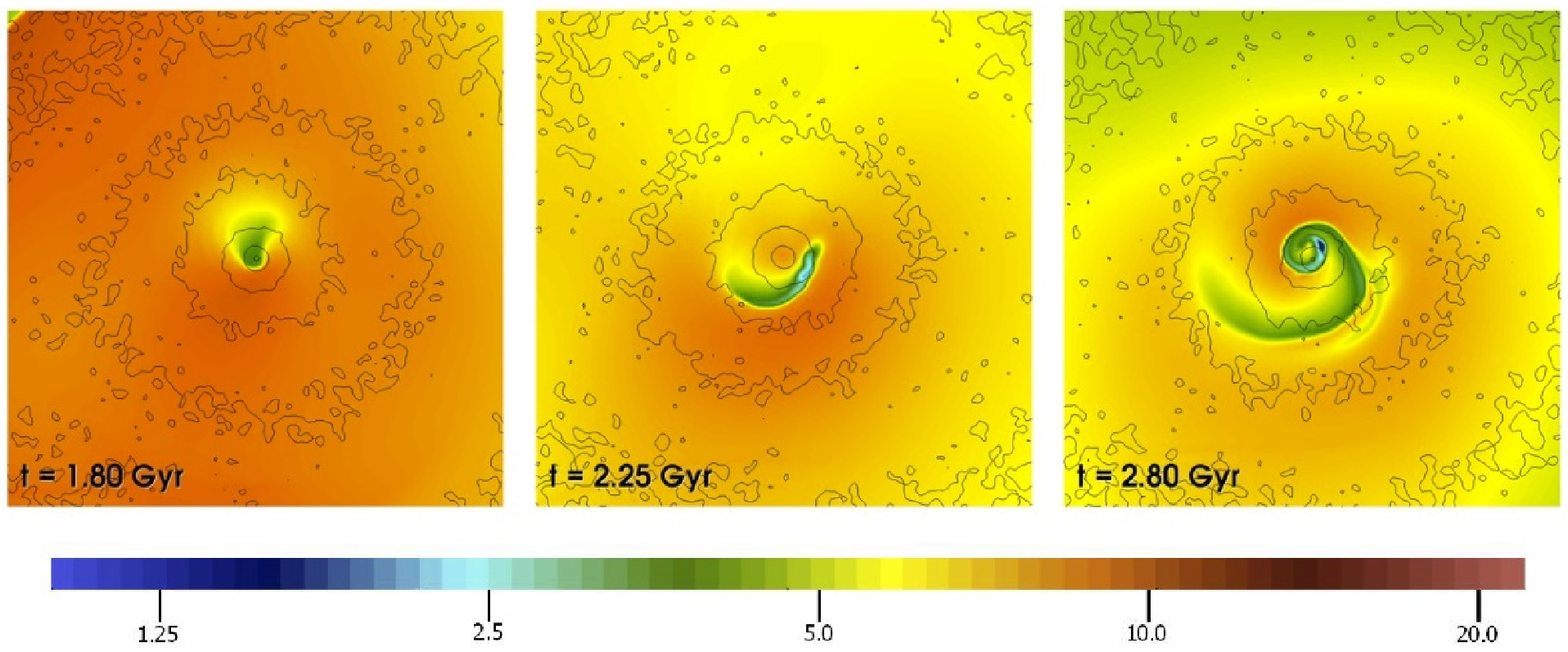}
\caption{Slices through the gas temperature in keV for simulation R5b500vc with dark matter contours overlaid. Each panel is 1 Mpc on a side.\label{fig:R5_b500_visc_cool}}
\end{center}
\end{figure*}

\begin{figure*}
\begin{center}
\plotone{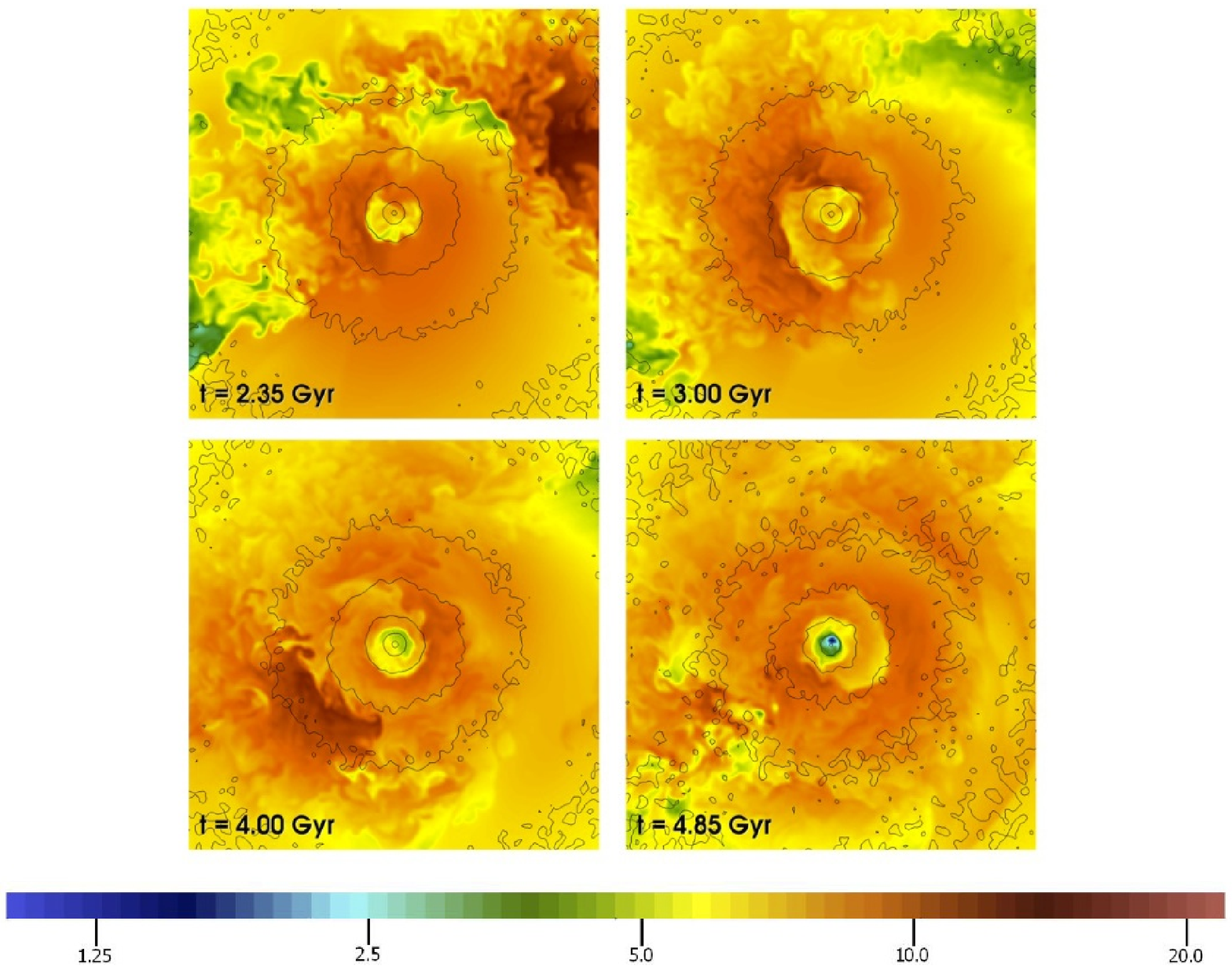}
\caption{Slices through the gas temperature in keV for simulation R20b200gc with dark matter contours overlaid. Each panel is 1 Mpc on a side.\label{fig:R20_b200_gas_cool}}
\end{center}
\end{figure*}

\begin{figure*}
\begin{center}
\plotone{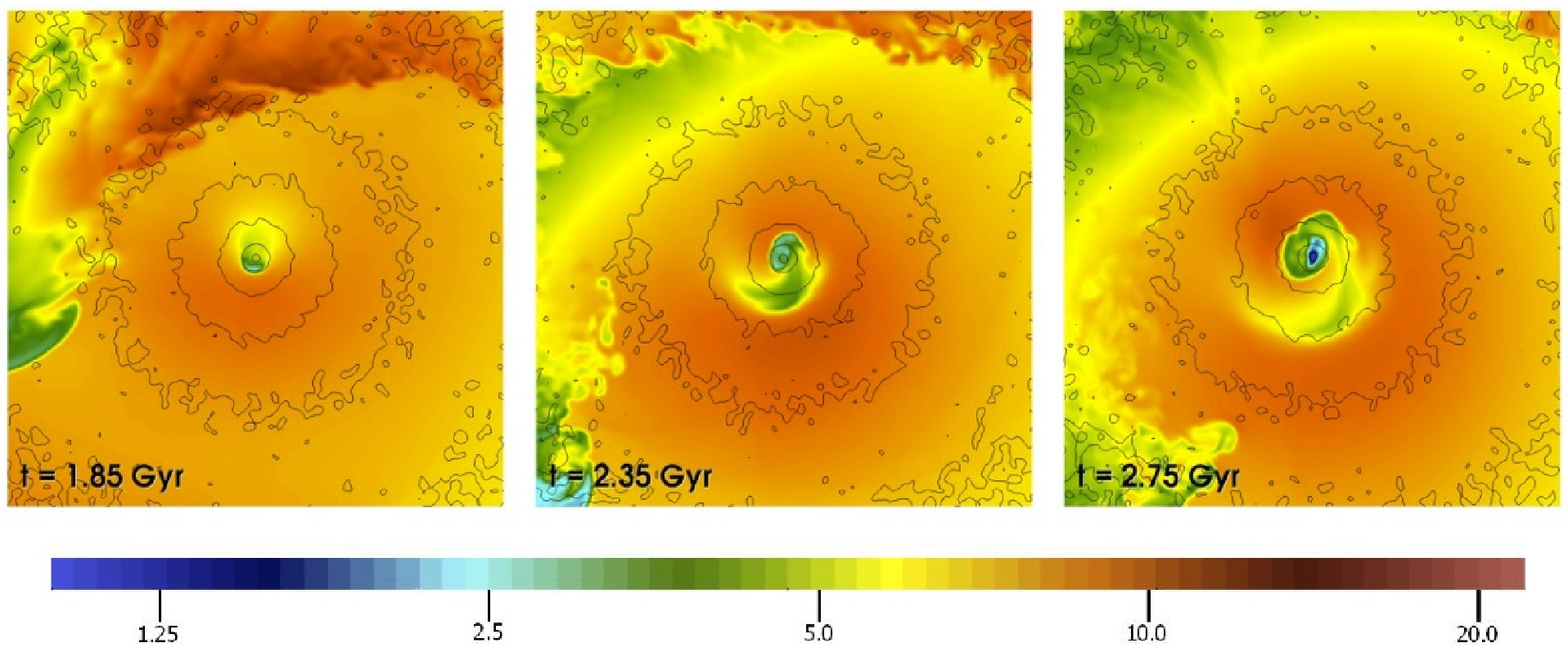}
\caption{Slices through the gas temperature in keV for simulation R20b1000gc with dark matter contours overlaid. Each panel is 1 Mpc on a side.\label{fig:R20_b1000_gas_cool}}
\end{center}
\end{figure*}

In the previous section, we showed that for a number of different configurations it is possible to generate an amount of core heating via sloshing which is at least comparable to the cooling rate, if not exceeding it, in the context of an adiabatic simulation. However, in a real cluster, cooling will also be modifying the state of the ICM, acting to lower the temperature of the gas and increase the density, making it more difficult for sloshing to effectively heat the core. Therefore, in order to make a direct determination of the effects of sloshing on the heating of the gas core and the possibility of quenching a cooling flow, we must self-consistently include the effects of gas cooling in our simulations. We have chosen four adiabatic simulations where sloshing will have a potentially interesting effect (R5b500, R5b500v, R20b200g, and R20b1000g) and have restarted them after the subcluster has passed (at the approximate moment where the core luminosity has returned to its initial value after the spike caused by the first subcluster passage), and switched on radiative cooling (simulations R5b500c, R5b500vc, R20b200gc, and R20b1000gc; the times when cooling has been switched on are given in Table \ref{tab:SimGrid}). In this way, we can gauge directly the effectiveness of the sloshing mechanism. Figures \ref{fig:R5_b500_cool} through \ref{fig:R20_b1000_gas_cool} show temperature slices through the center of the domain with dark matter density contours overlaid, analogous to Figures \ref{fig:R5_b500} to \ref{fig:R20_b1000_gas}. Sloshing occurs in each of these simulations in much the same way as their adiabatic counterparts. The major difference is that in each of these cases, at a time earlier than the end of the correspoding adiabatic simulation, a runaway cooling flow develops within the central $\sim$10-20~kpc. In each case, the simulation is stopped when the central cooling time reaches a value $t_c < 1$~Myr, since after this point the necessary numerical timestep becomes prohibitively small.

\begin{figure}
\begin{center}
\plotone{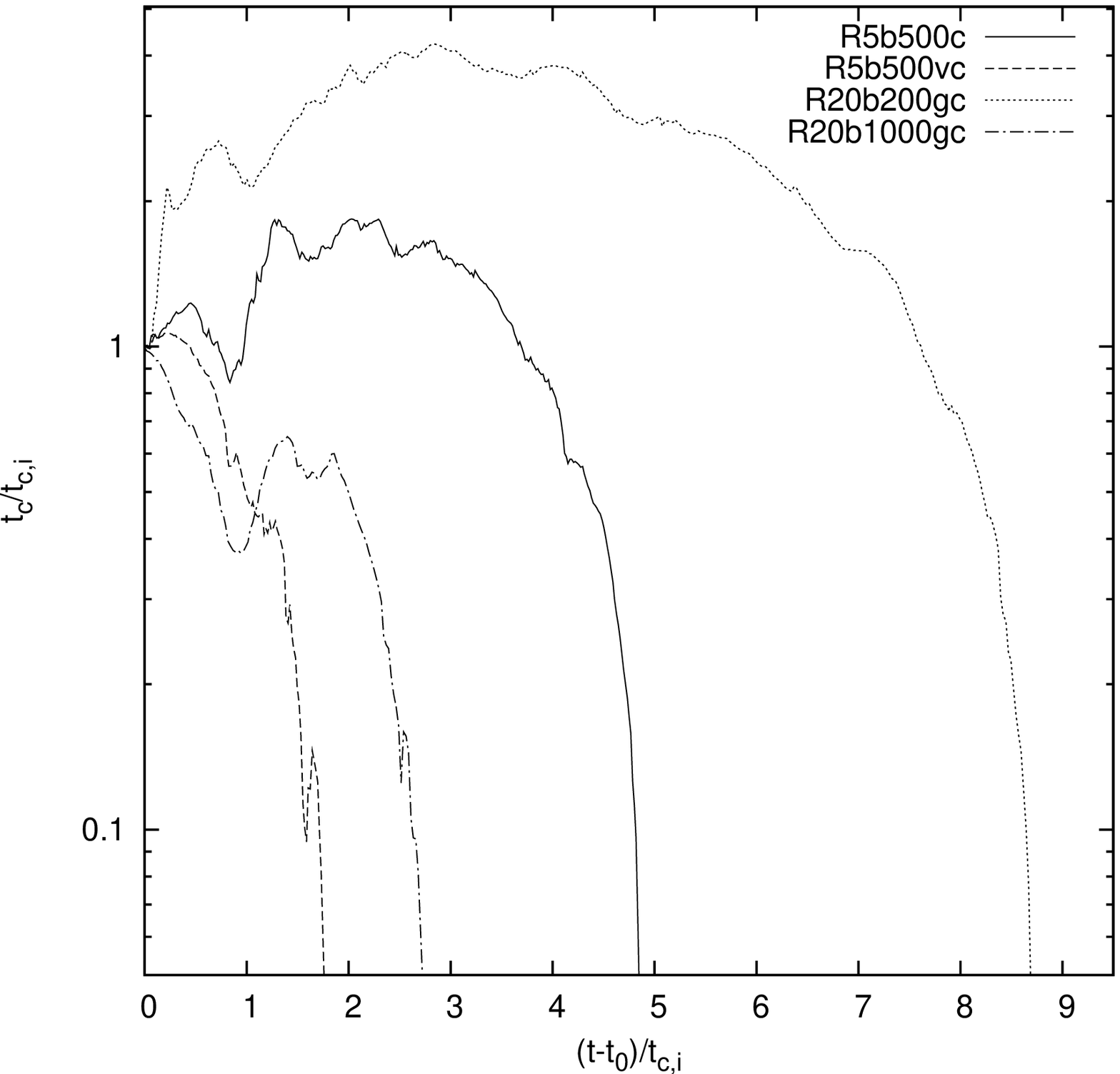}
\caption{Cooling time at the center of the cluster vs. simulation time for the cooling simulations. The simulation time is scaled to the epoch when cooling begins.\label{fig:cooling_time}}
\end{center}
\end{figure}

\begin{figure*}
\begin{center}
\includegraphics[width=0.45\textwidth]{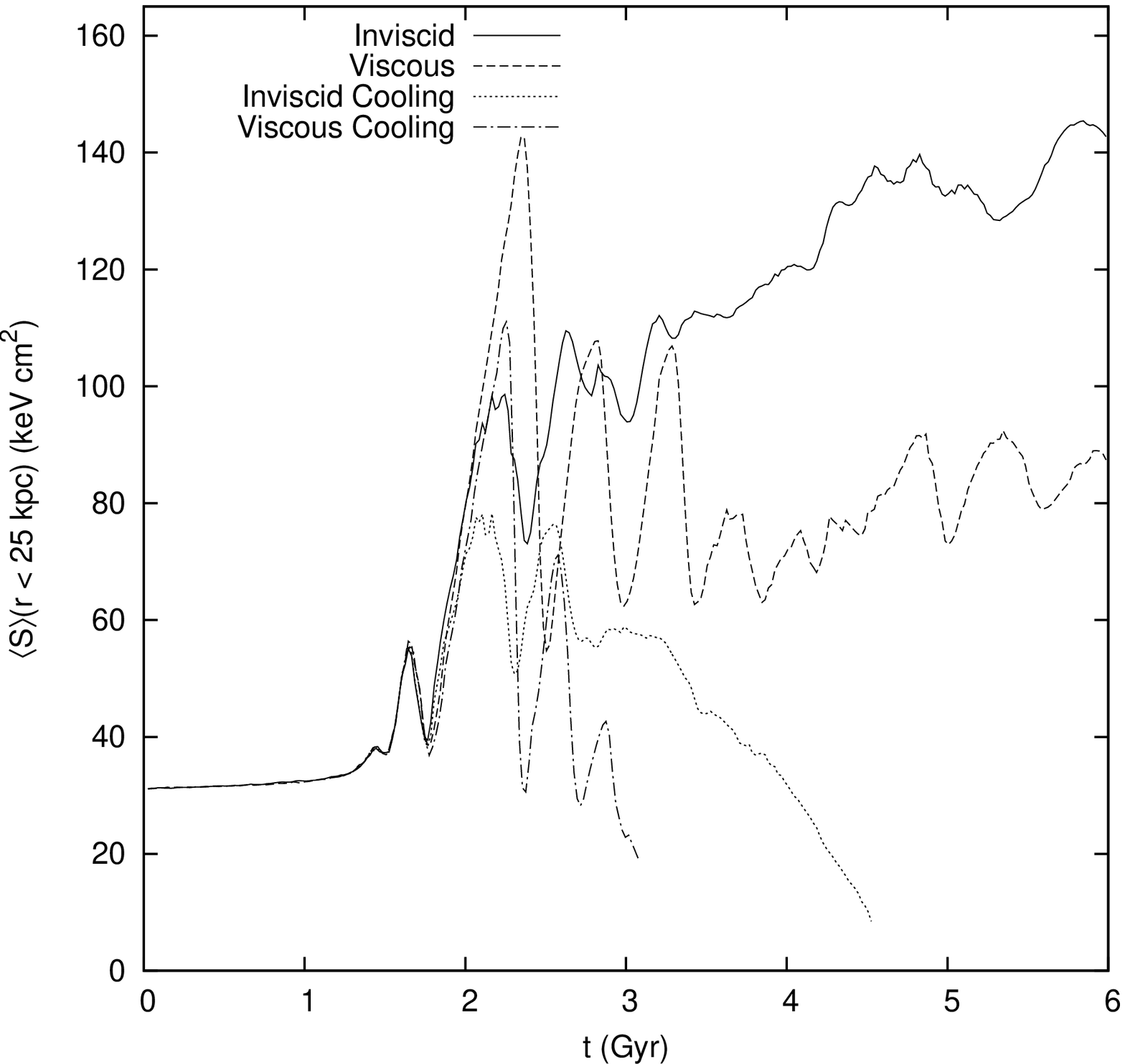}
\includegraphics[width=0.45\textwidth]{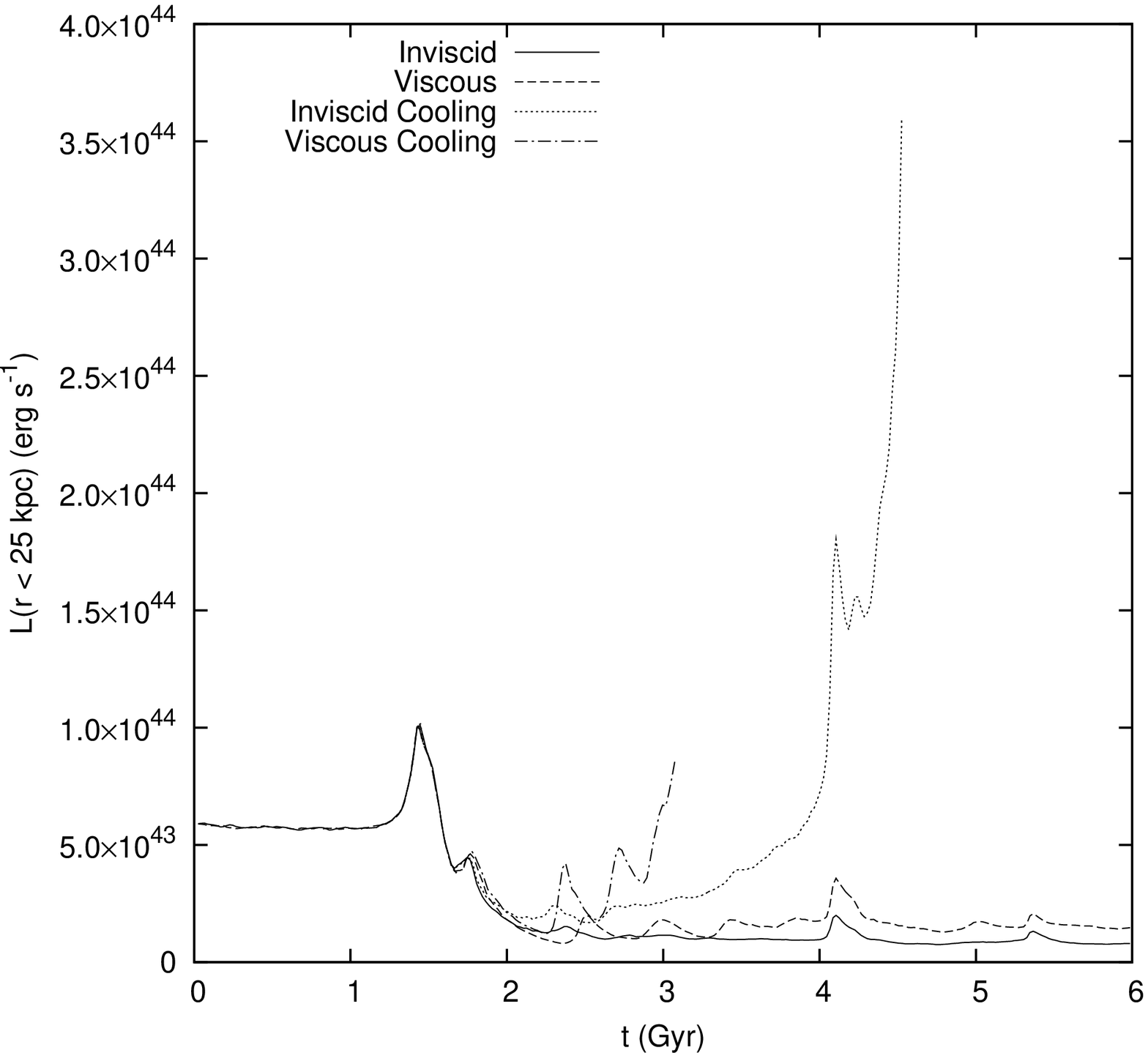}
\par
\includegraphics[width=0.45\textwidth]{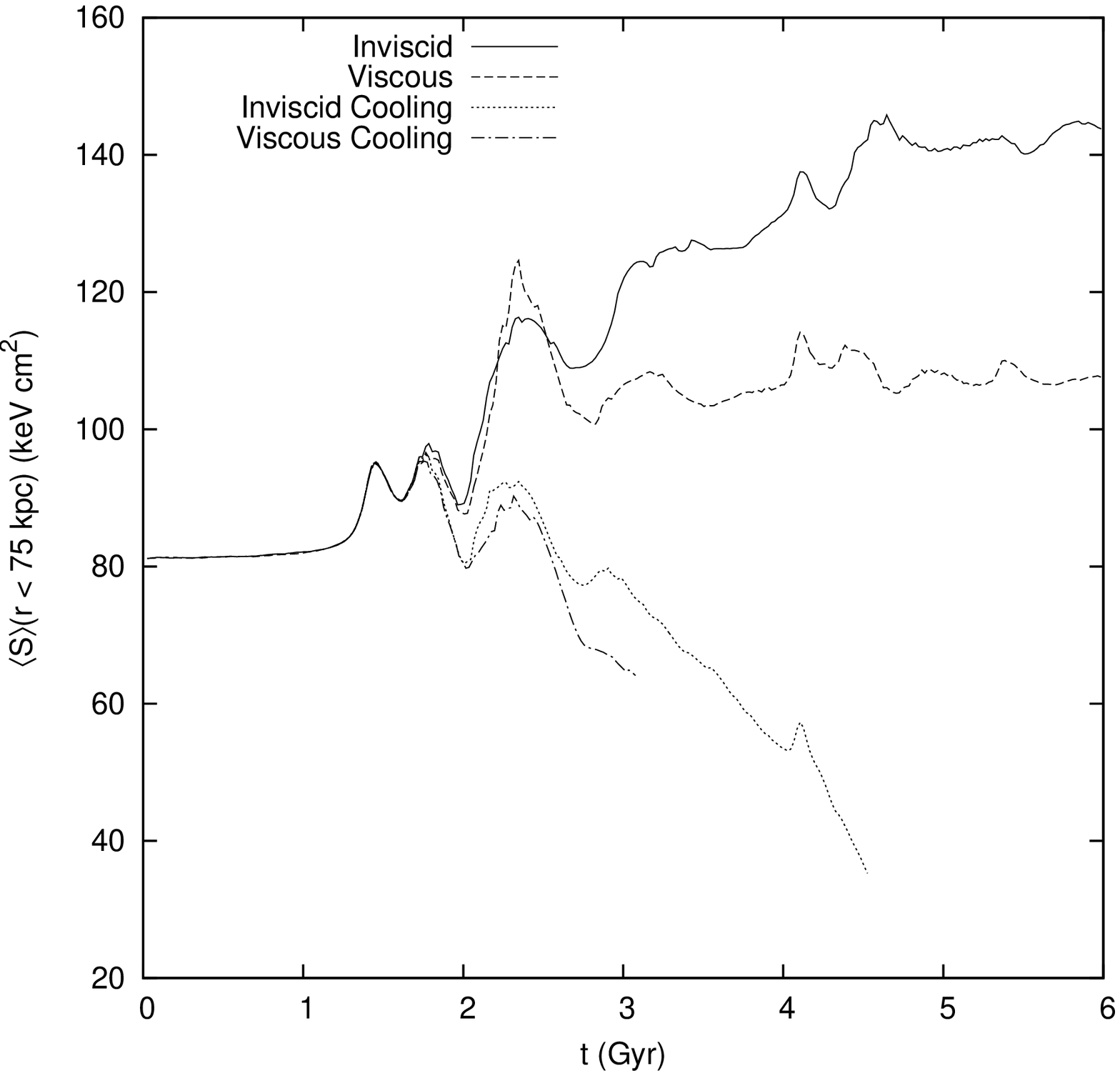}
\includegraphics[width=0.45\textwidth]{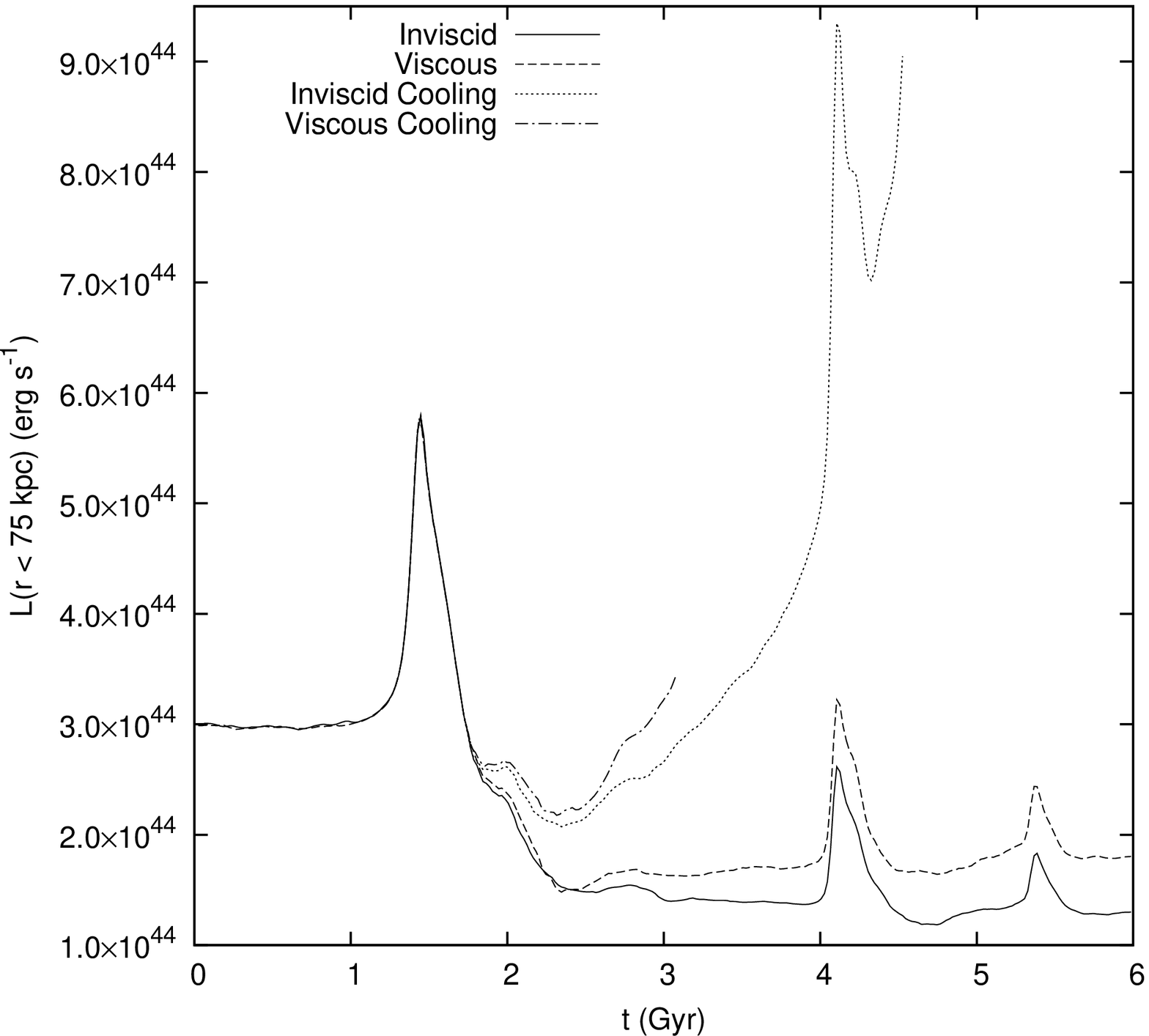}
\caption{Evolution of the $R$ = 5, $b$ = 500 kpc simulations. Left: Average entropy vs. simulation time. Right: Luminosity vs. simulation time.\label{fig:compare1}}
\end{center}
\end{figure*}

\begin{figure*}
\begin{center}
\includegraphics[width=0.45\textwidth]{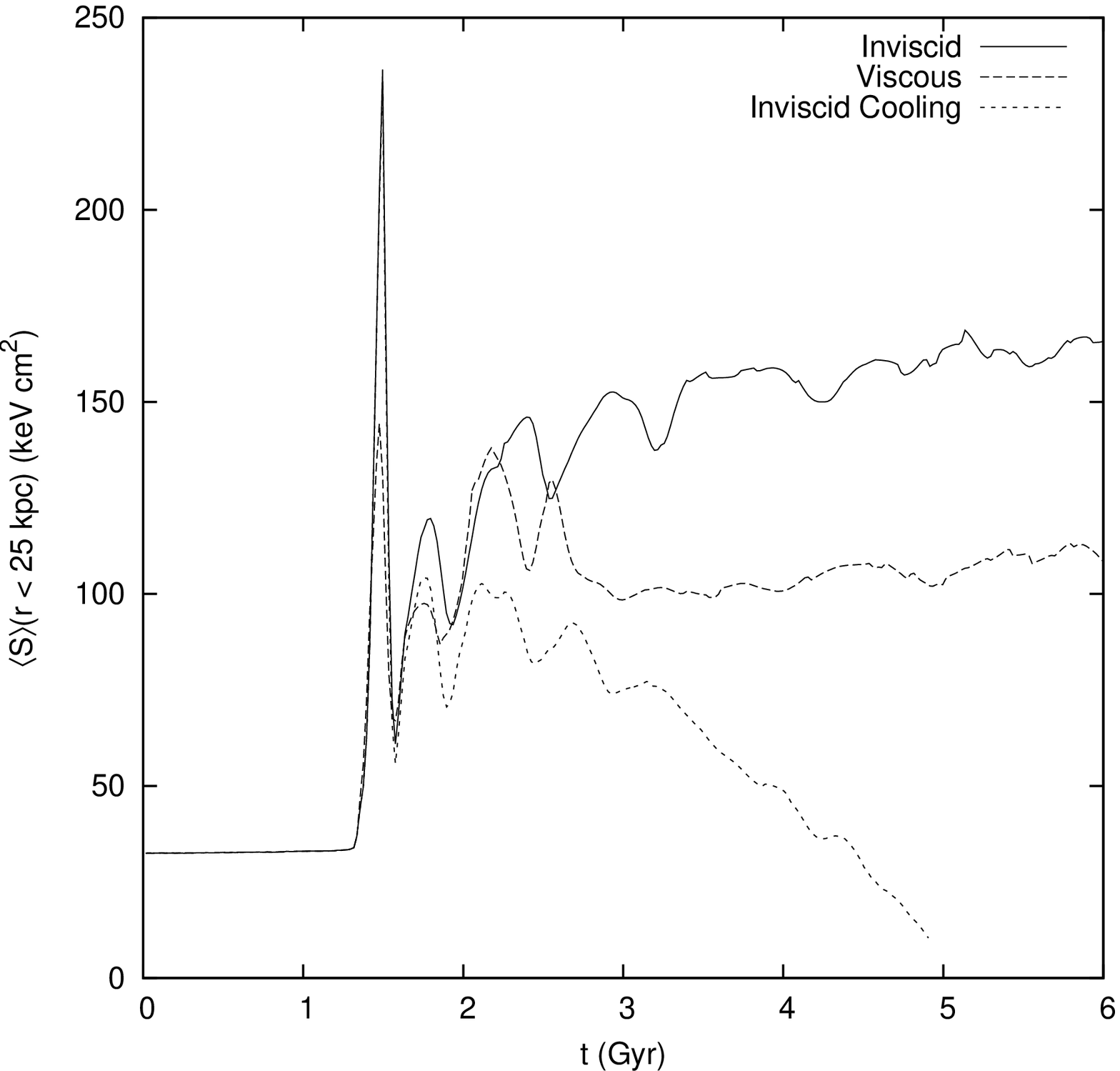}
\includegraphics[width=0.45\textwidth]{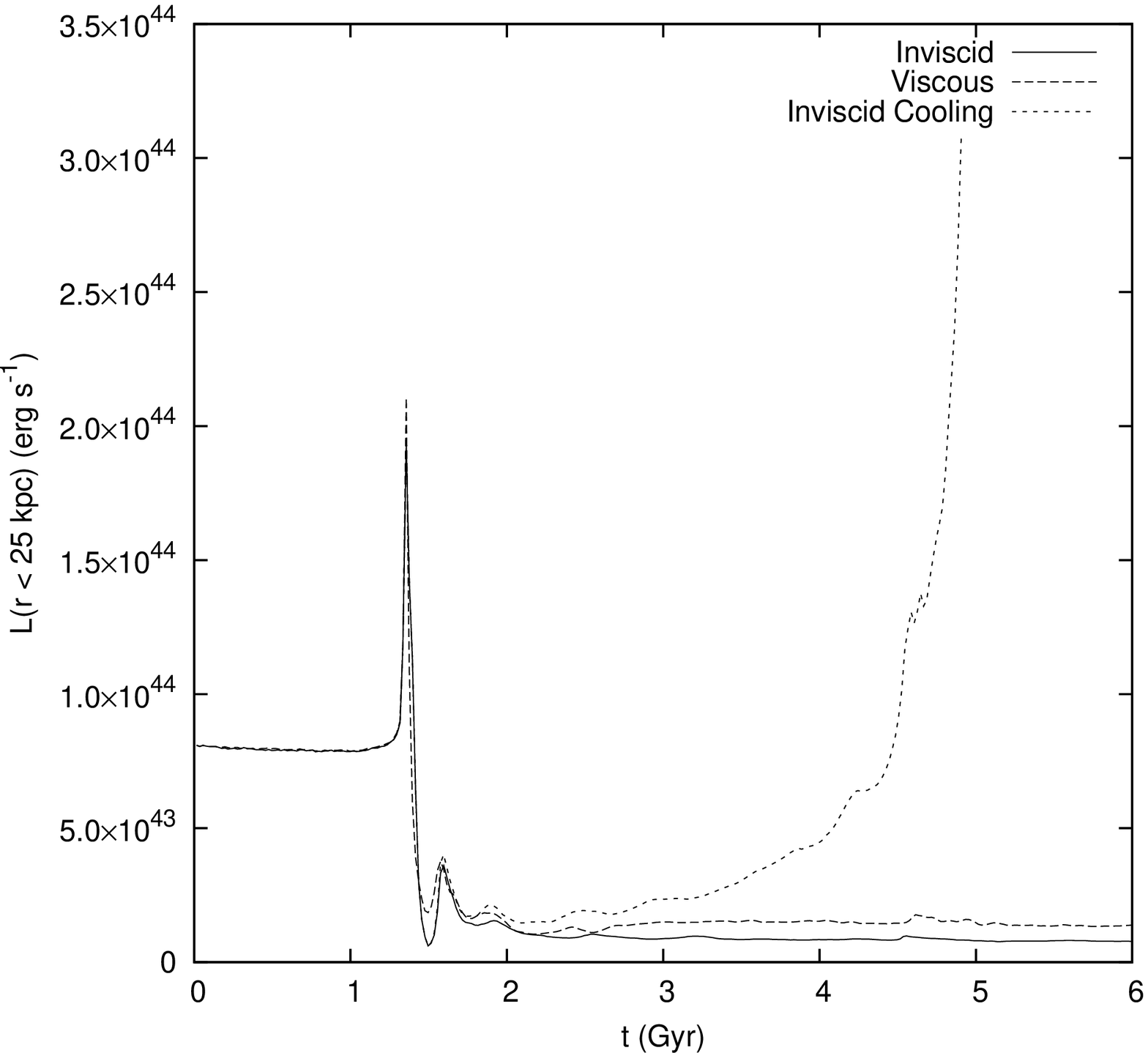}
\par
\includegraphics[width=0.45\textwidth]{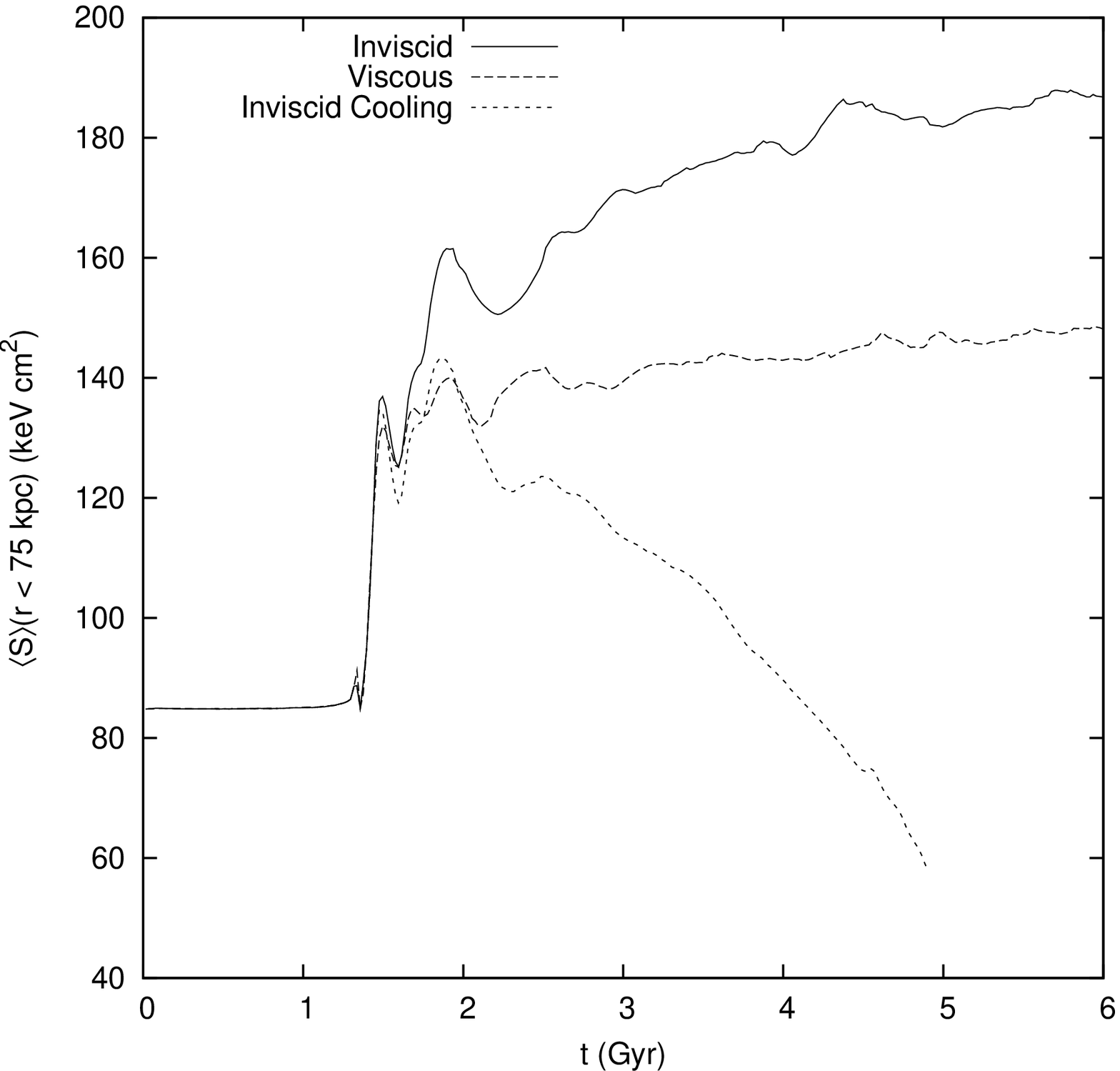}
\includegraphics[width=0.45\textwidth]{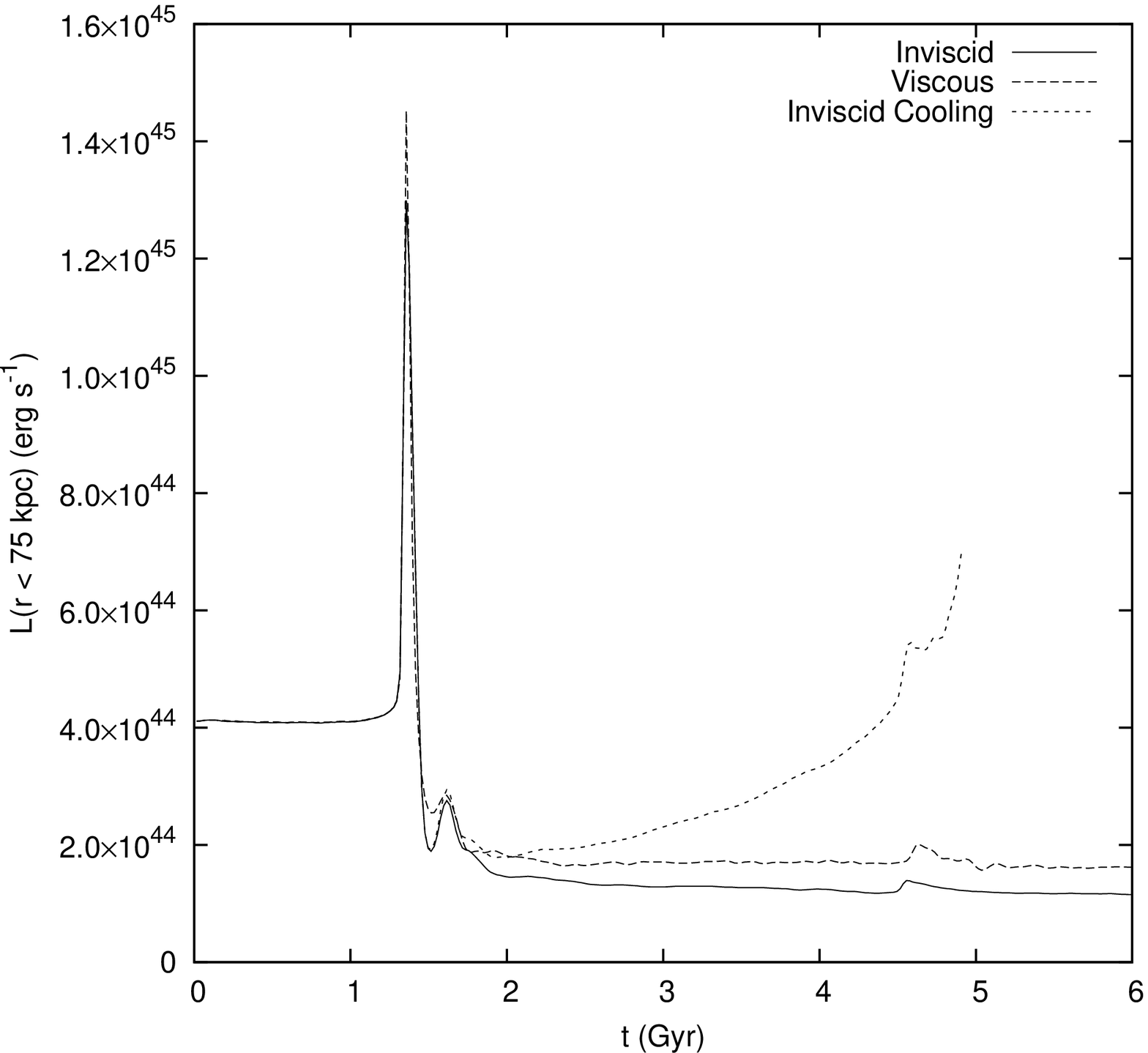}
\caption{Evolution of the $R$ = 20, $b$ = 200 kpc with gas simulations. Left: Average entropy vs. simulation time. Right: Luminosity vs. simulation time.\label{fig:compare2}}
\end{center}
\end{figure*}

\begin{figure*}
\begin{center}
\includegraphics[width=0.45\textwidth]{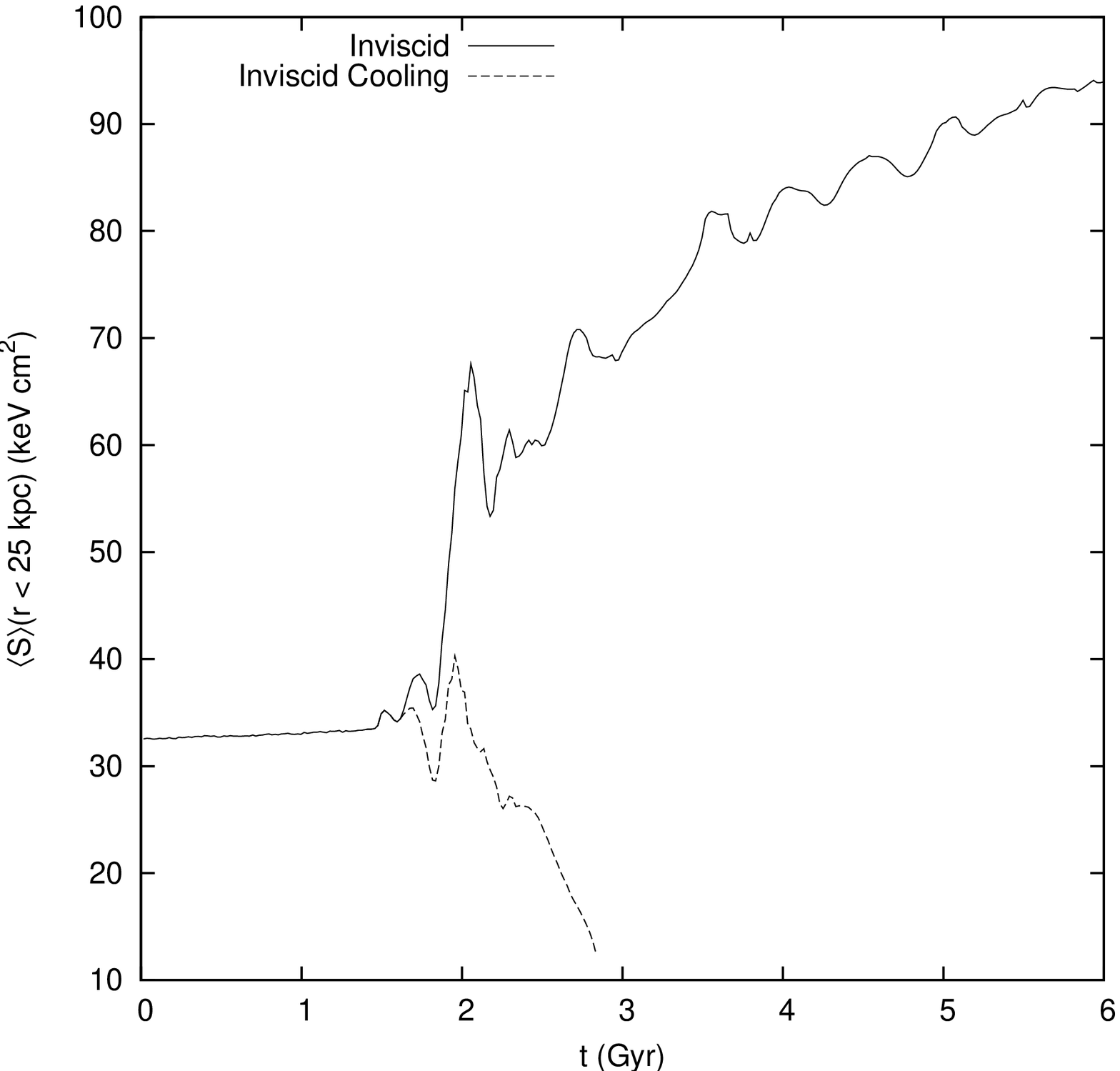}
\includegraphics[width=0.45\textwidth]{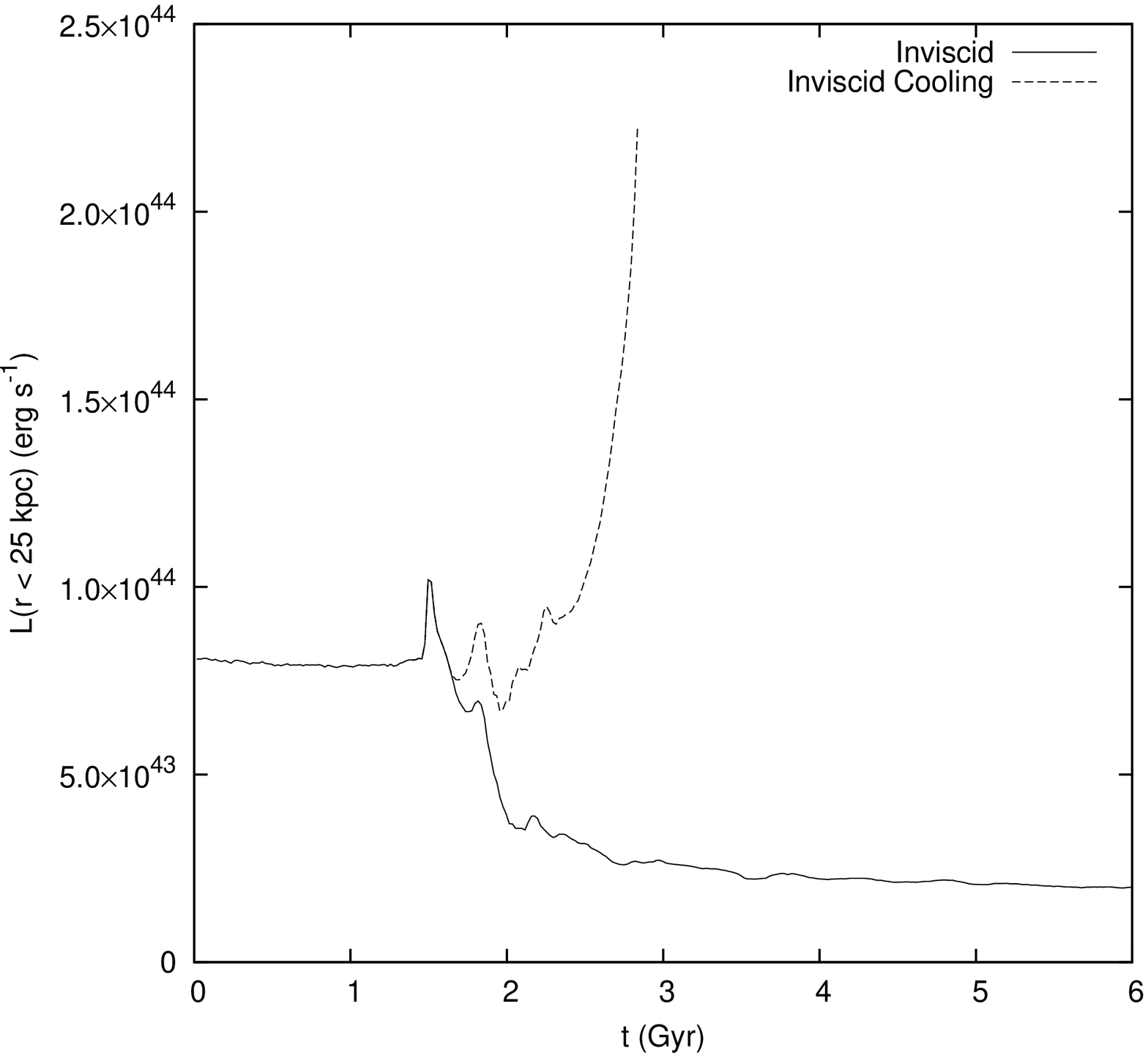}
\par
\includegraphics[width=0.45\textwidth]{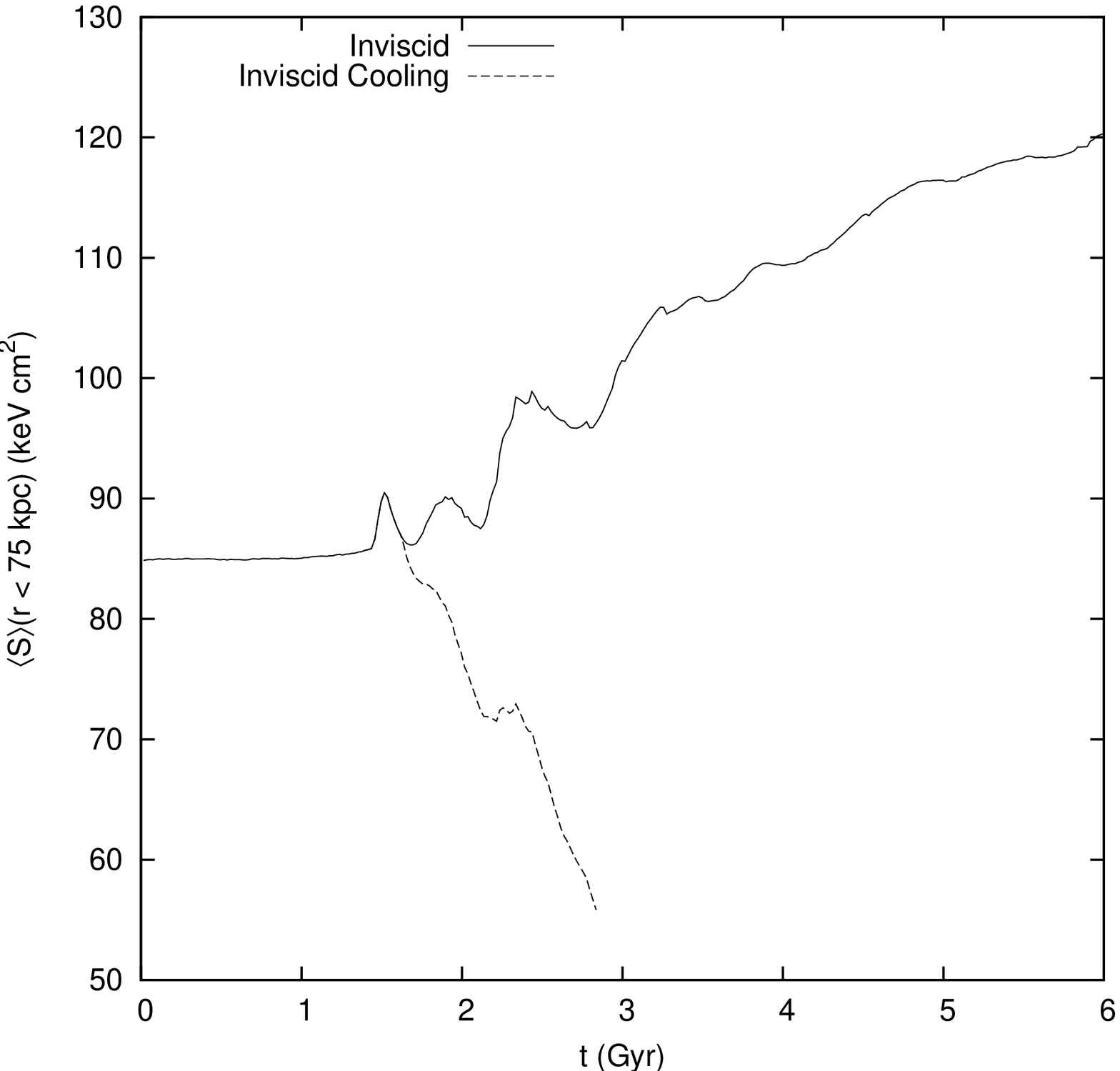}
\includegraphics[width=0.45\textwidth]{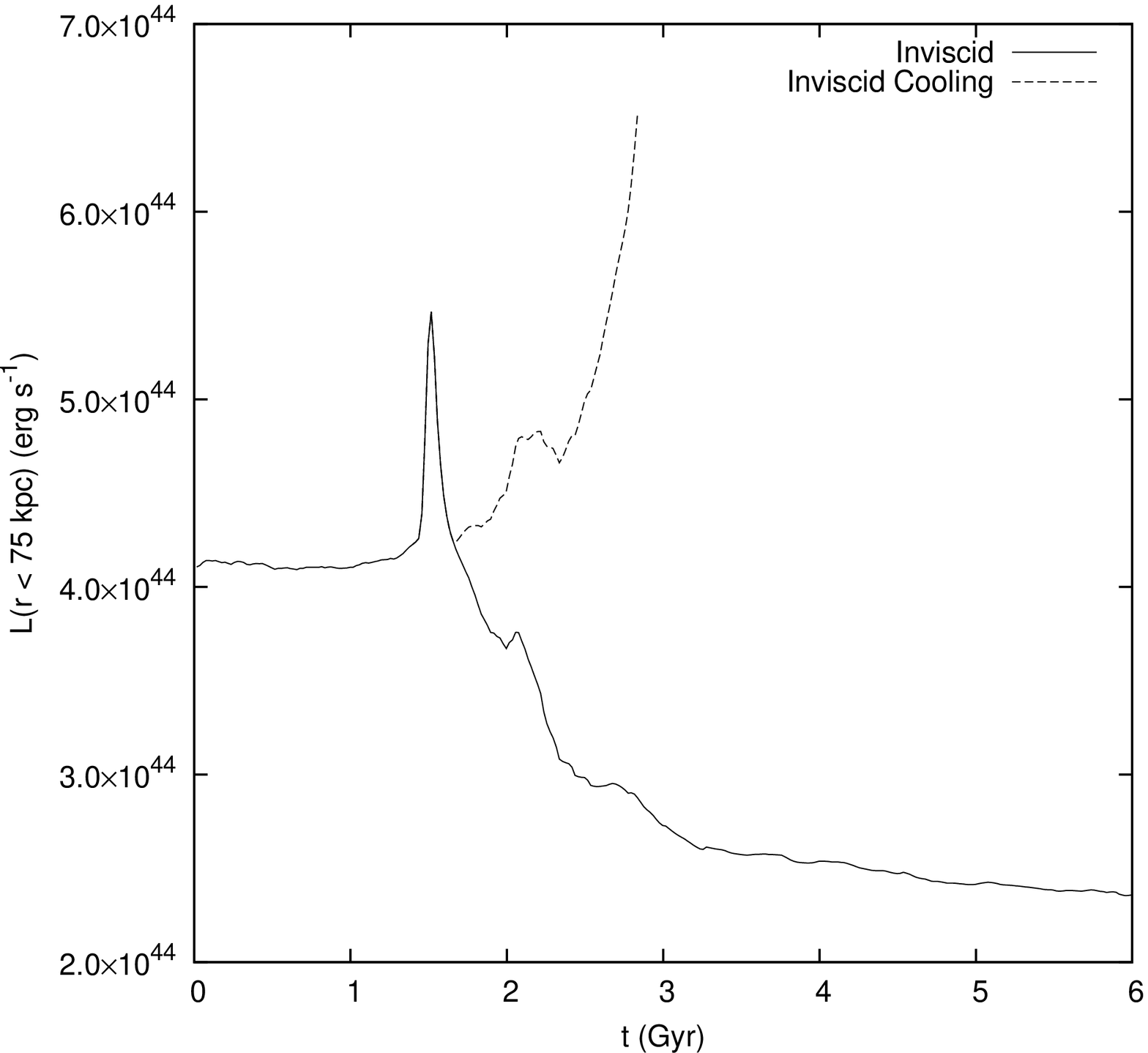}
\caption{Evolution of the $R$ = 20, $b$ = 1000 kpc with gas simulations. Left: Average entropy vs. simulation time. Right: Luminosity vs. simulation time.\label{fig:compare3}}
\end{center}
\end{figure*}

\begin{figure}
\begin{center}
\includegraphics[width=0.45\textwidth]{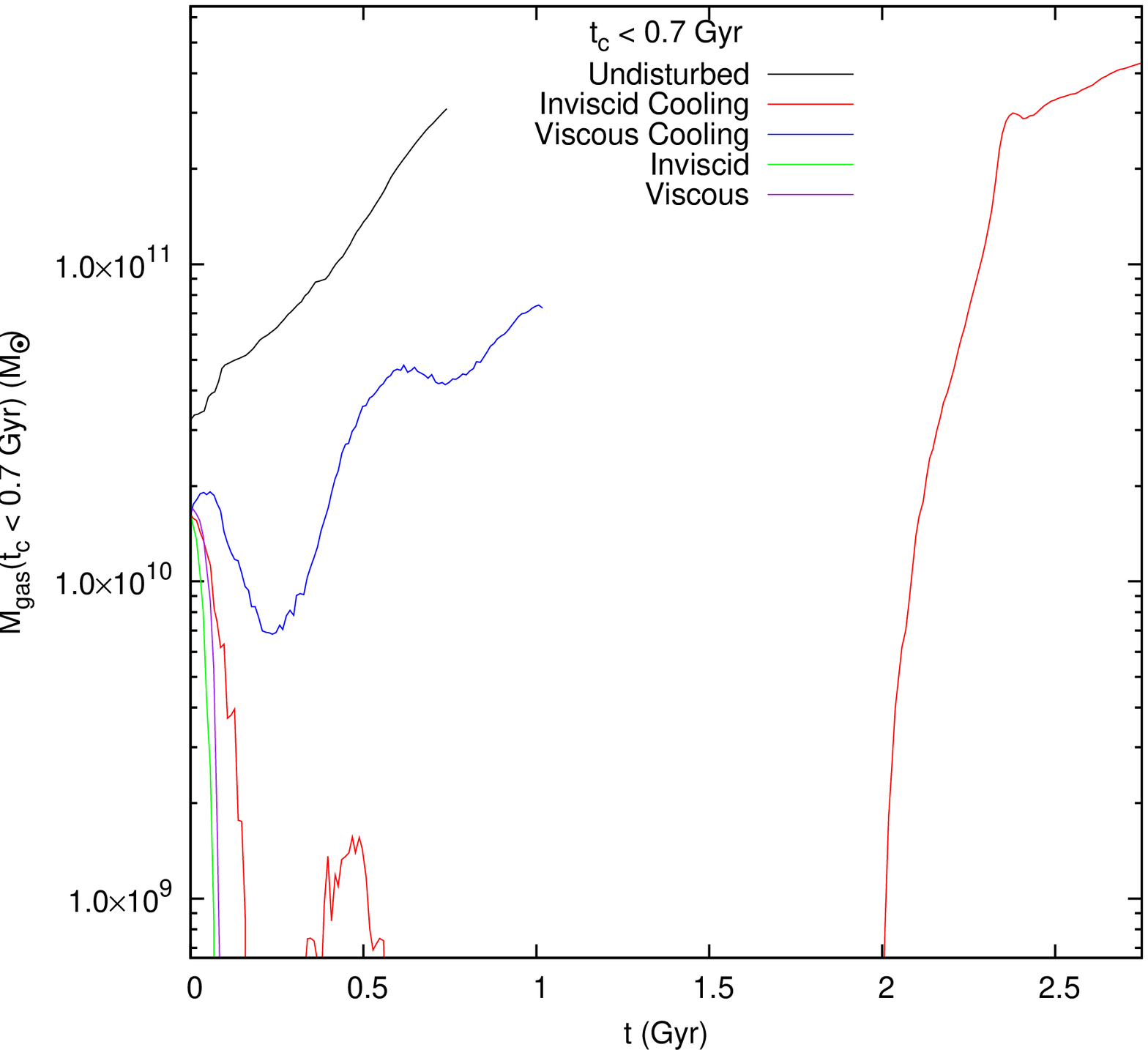}
\includegraphics[width=0.45\textwidth]{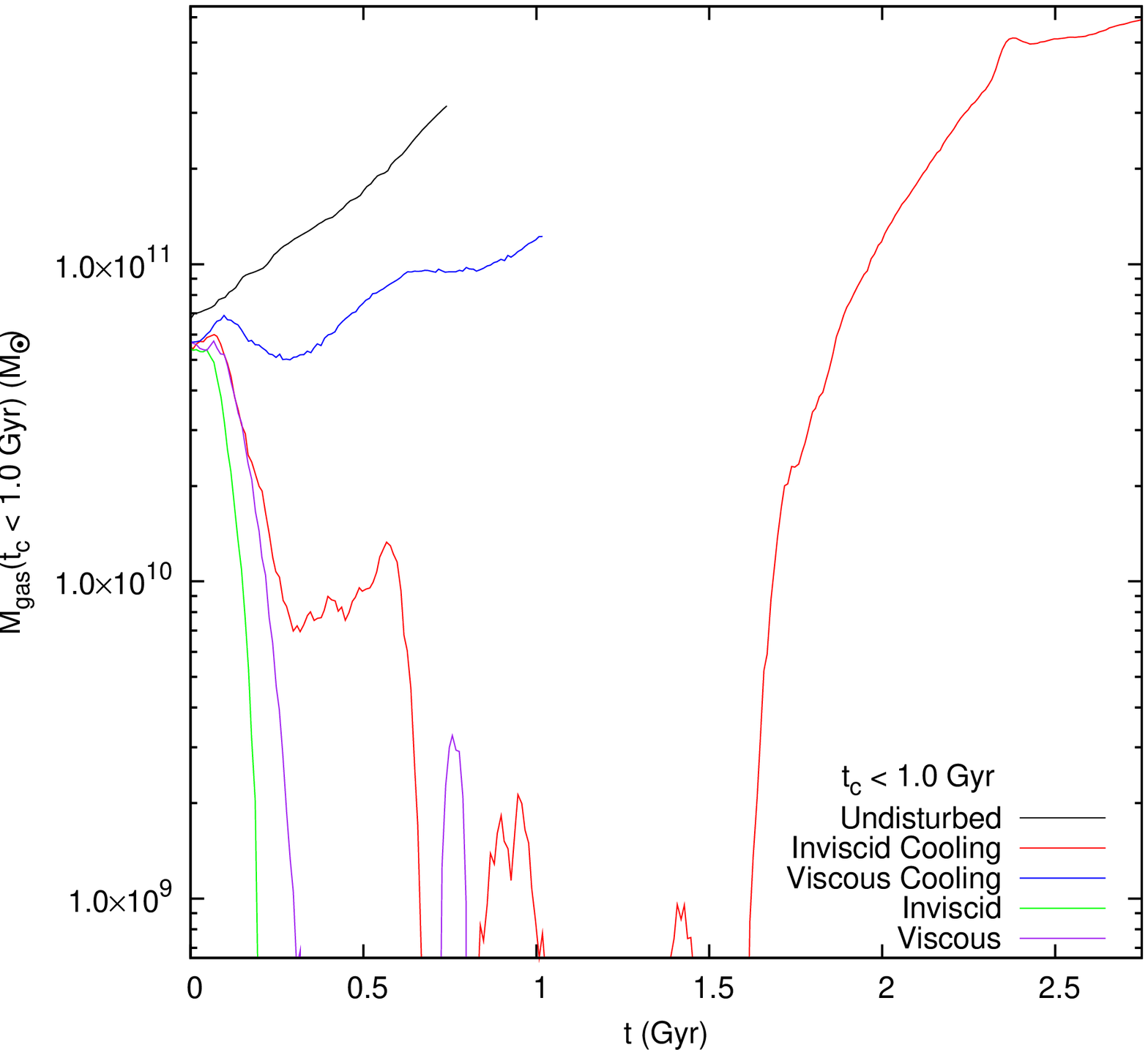}
\par
\includegraphics[width=0.45\textwidth]{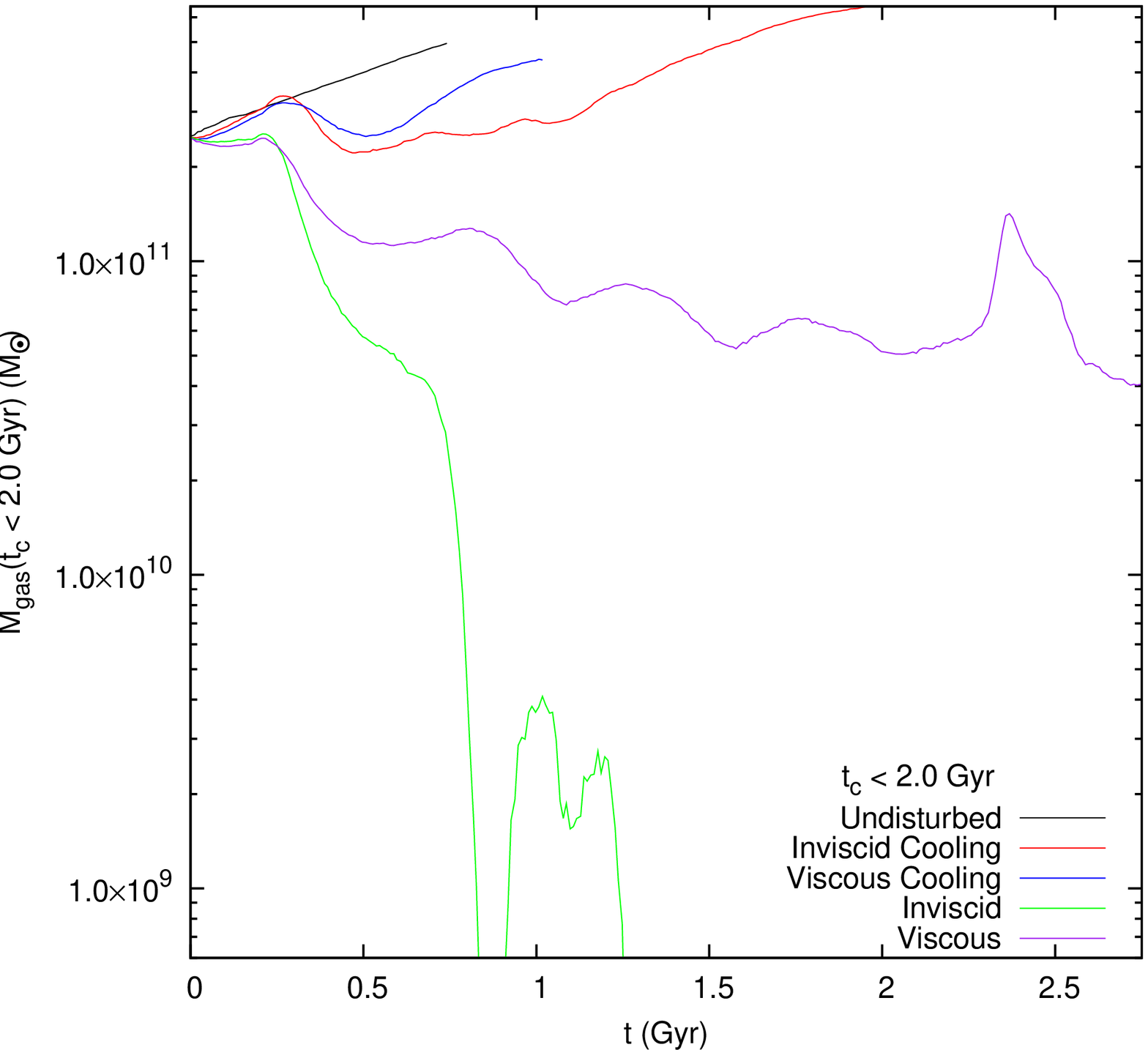}
\caption{Evolution of the gas mass with a cooling time lower than a given value for simulations with $R$ = 5 and $b$ = 500~kpc. Left: Gas mass with cooling time $t_c < 0.7$~Gyr. Center: Gas mass with cooling time $t_c < 1.0$~Gyr. Right: Gas mass with cooling time $t_c < 2.0$~Gyr.\label{fig:mcoolR5}}
\end{center}
\end{figure}

To varying degrees, depending on the parameter set, sloshing is able to provide significant energy input over timescales considerably longer than the merger crossing timescale. Figure \ref{fig:cooling_time} shows the cooling time in the central resolution element in each cooling simulation vs. simulation time, both measured in units of the cooling time at the moment when cooling was ``switched on'' (the former being $t_{c,i}$, the latter $t_0$), which in these simulations is $\approx$500~Myr. Each of the simulations is effective at staving off a catastrophe for an interval of time, but there is a range of effectiveness. In general, as expected, the cooling simulations that prevent a catastrophe for a longer interval of time roughly correspond to the adiabatic simulations that have a higher ratio of heating to cooling (see Table \ref{tab:heat_to_cool}). For example, simulation R5b500c is seen to be able to prevent a cooling catastrophe near the cluster center for 5 initial cooling times, whereas simulation R5b500vc is only able to do so for 2.5 initial cooling times. The ratio of heating to cooling is higher in the corresponding simulation R5b500 than it is in the simulation R5b500v. This is not a ``hard-and-fast'' rule, however, as simulation R20b1000gc staves off a catastrophe for half an initial cooling time longer than simulation R5b500vc, even though simulation R5b500v has a higher rate of heating to cooling than simulation R20b1000g. This has to do with the smaller amount of mixing in simulation R5b500vc than in simulation R20b1000gc due to the action of viscosity in the former simulation (see Section \ref{sec:viscosity}). 

Figures \ref{fig:compare1} through \ref{fig:compare3} show the evolution of average entropy (within radii of $r < 25, 75$ kpc) and luminosity (within radii of $r < 25, 75$ kpc) for all the simulations. Simulations with viscosity and with cooling have been grouped together with their respective adiabatic and inviscid counterparts for comparison. In the adiabatic simulations, the average entropy goes up and the luminosity goes down. The opposite effects occur in the cooling simulations. However, the simulations with higher degrees of sloshing (simulations with stronger disturbances, subclusters with gas, and an inviscid ICM) are able to maintain entropy and luminosity close to the original values for a longer interval of time. 

\begin{figure}
\begin{center}
\includegraphics[width=0.45\textwidth]{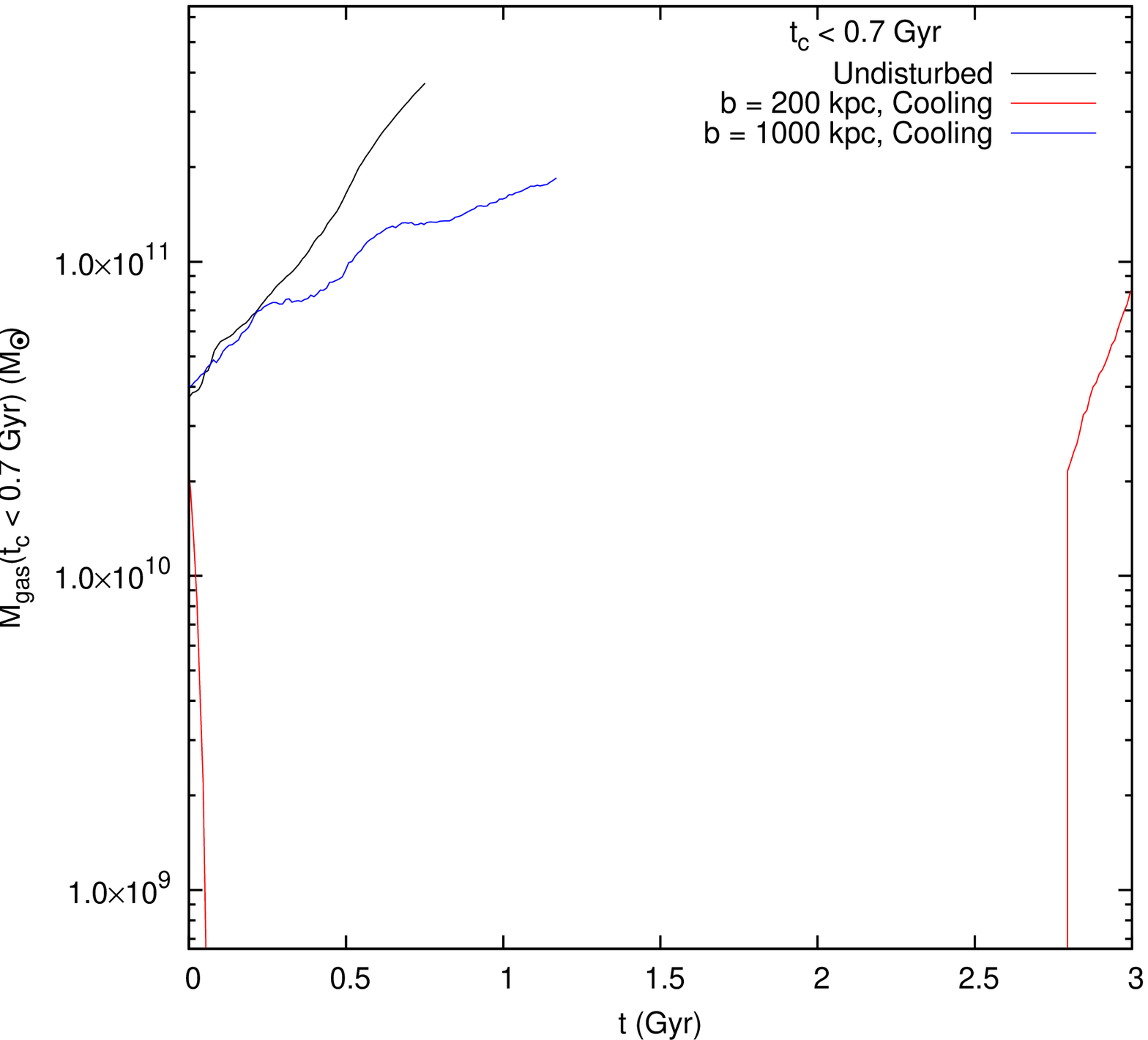}
\includegraphics[width=0.45\textwidth]{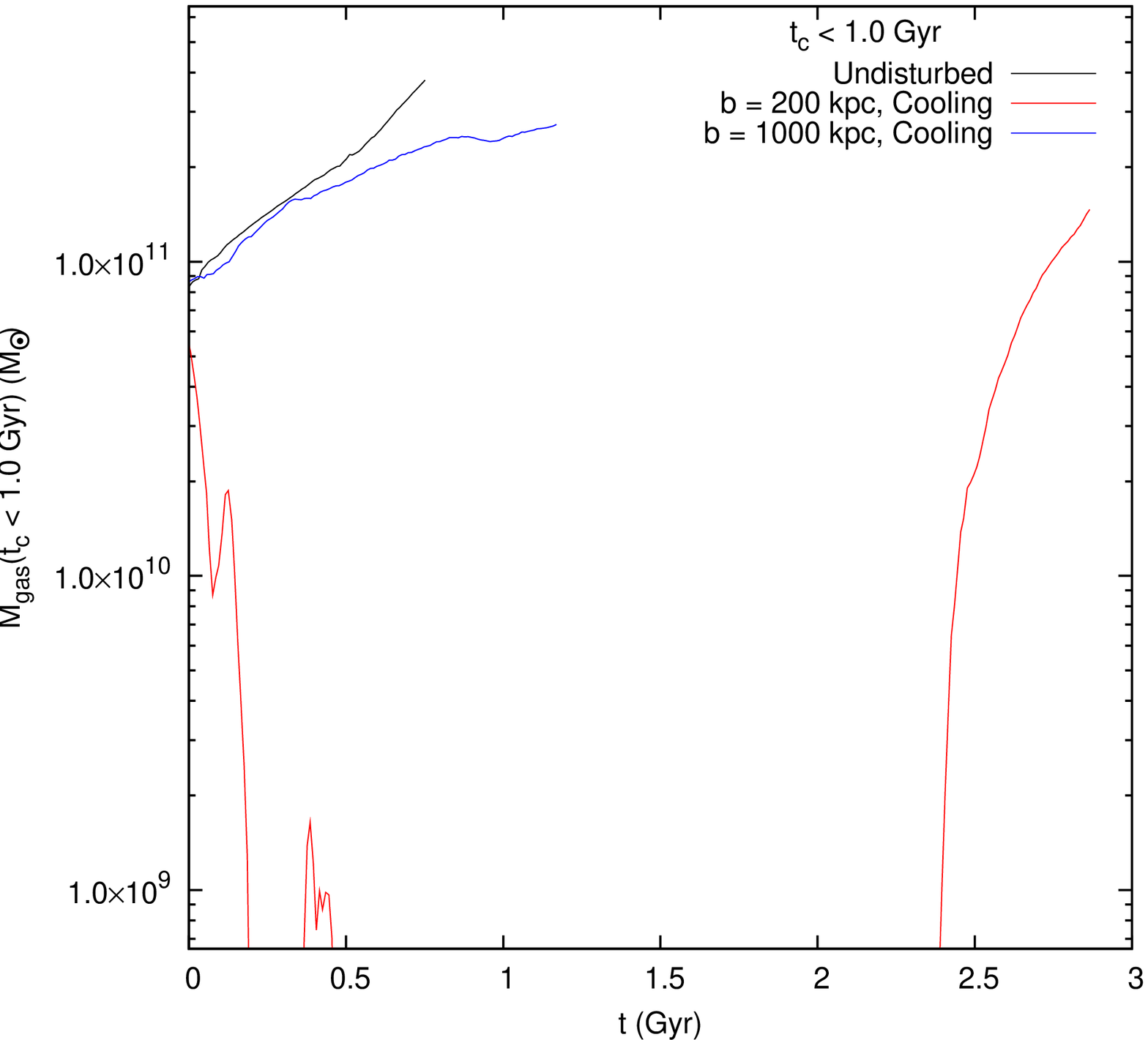}
\par
\includegraphics[width=0.45\textwidth]{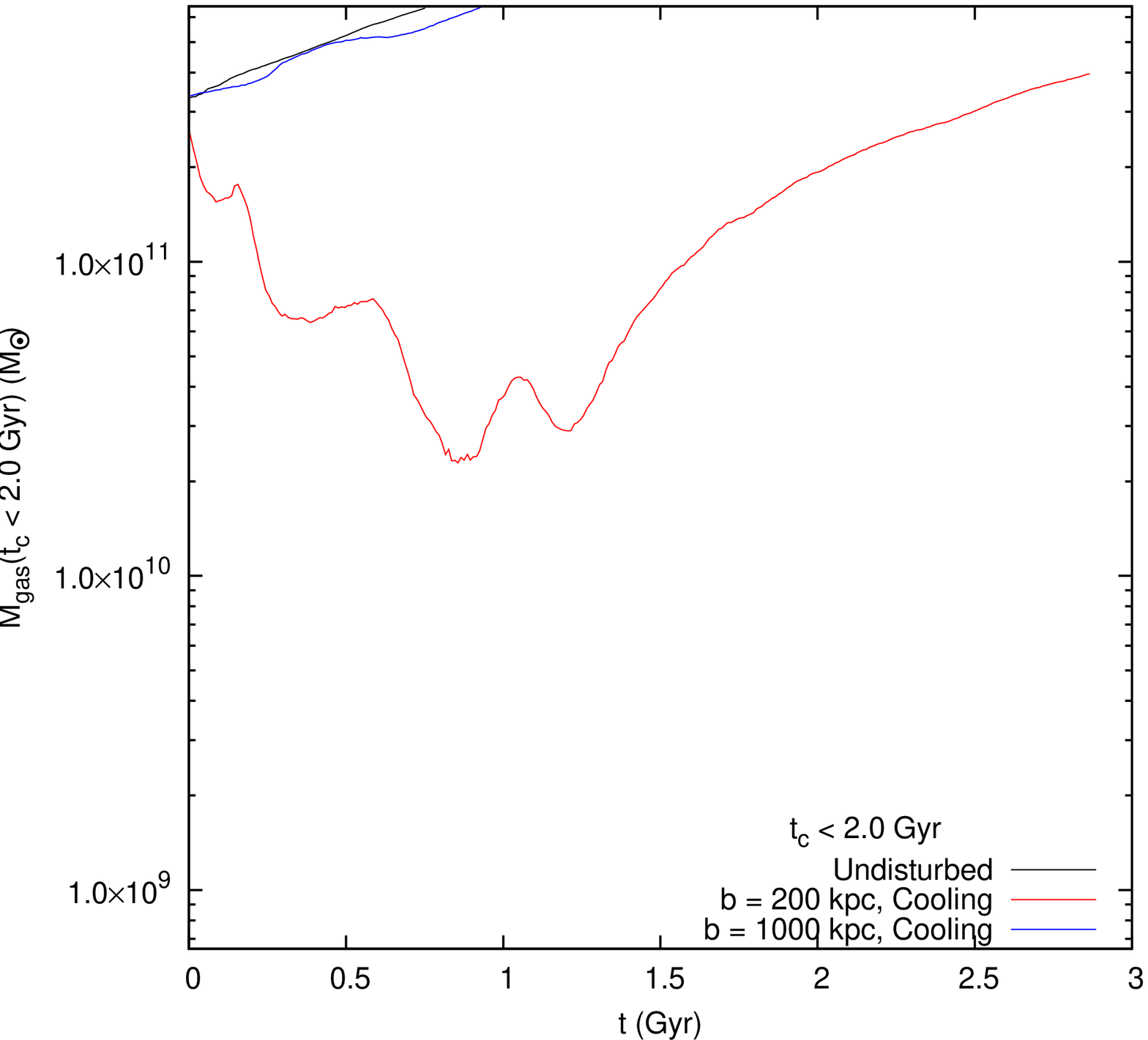}
\caption{Evolution of the gas mass with a cooling time lower than a given value for simulations with $R$ = 20 and subclusters with gas. Left: Gas mass with cooling time $t_c < 0.7$~Gyr. Center: Gas mass with cooling time $t_c < 1.0$~Gyr. Right: Gas mass with cooling time $t_c < 2.0$~Gyr.\label{fig:mcoolR20}}
\end{center}
\end{figure}

A way of gauging the effectiveness of sloshing that is perhaps more relevant for the bulk of the cooling core is to determine the amount of gas that is cooling above a certain rate, or, equivalently, that has a cooling time less than a certain value. In an undisturbed cluster with no sources of heating, the mass of gas that has a cooling time less than a certain value should increase unabated very quickly. If sloshing is effective, it should be able to stave this increase off for an interval of time. Figures \ref{fig:mcoolR5} and \ref{fig:mcoolR20} show the gas mass with a cooling time less than $t_c < 0.7$, 1.0, and 2.0~Gyr for the cooling simulations, grouped into the $R$ = 5, $b$ = 500~kpc cases and the $R$=20, gas-filled subcluster cases (in the former, the non-cooling curves are also shown for comparison). A corresponding curve for the initial subcluster is also plotted for comparison (since the main cluster has undergone the initial portion of the evolution in the merging simulations, the initial mass of this gas is not precisely the same as in the undisturbed cluster, but it is similar enough for our purposes). For the undisturbed cluster, the mass of gas with a low cooling time continuously increases, within less than 1~Gyr reaching $\sim$6 times the original mass of gas with a cooling time less than 0.7~Gyr and reaching $\sim$2 times the original mass of gas with a cooling time less than 2.0~Gyr. In each of the cases with sloshing, this increase of gas mass is slowed down or even halted for an interval of time, ranging from 0.25-1.0~Gyr in the weakest cases (simulations R5b500vc and R20b1000gc) to 2.0-3.0~Gyr in the stronger cases (simulations R5b500c and R20b200gc).

\section{Discussion\label{sec:disc}}

\subsection{The Effectiveness of Sloshing\label{sec:effectiveness}}

The adiabatic simulations demonstrated that sloshing can provide a significant amount of heat to the cluster core. In each case, sloshing brought gas of high entropy from larger radii into contact with lower-entropy gas from the cluster core, resulting in a net increase in entropy in the core. In nearly every case, the central entropy of the merger remnant is increased significantly from its original value (see Figure \ref{fig:e_profiles}). Additionally, the effect of this raising of the core entropy is to decrease the efficiency of the cooling of the core due to the associated expansion of the gas. 

The simulations with cooling demonstrate that sloshing is a viable mechanism to stave off a cooling flow. The heat produced by sloshing was able to suppress the increase of the mass of gas with low cooling times, for a length of time up to a few Gyr in the case of strong disturbances (such as simulations R5b500c and R20b200gc). Within this interval of time, the cluster is likely to have successive encounters with other subclusters, which will help to sustain the sloshing of the gas and therefore continue to provide a source of heat to offset cooling. 

Obviously, sloshing will coexist with other sources of heating present in the ICM. After the sloshing has subsided, gas cooling will at some point reestablish a high-density, low-temperature gas core in the absence of another period of sloshing caused by another merger (or a secondary passage of the original subcluster), or some alternative source of heating. However, if a source of feedback (e.g. AGN) already exists, the softening of the gas core created by sloshing will decrease the cooling rate and hence reduce the need for energy input from other mechanisms. Sloshing may then work in tandem with other sources of feedback to prevent a cooling catastrophe. Investigating this possibility in detail will require simulation studies beyond the scope of this work. 

\subsection{The Effect of An Explicit Viscosity\label{sec:viscosity}}

\begin{figure*}
\begin{center}
\plotone{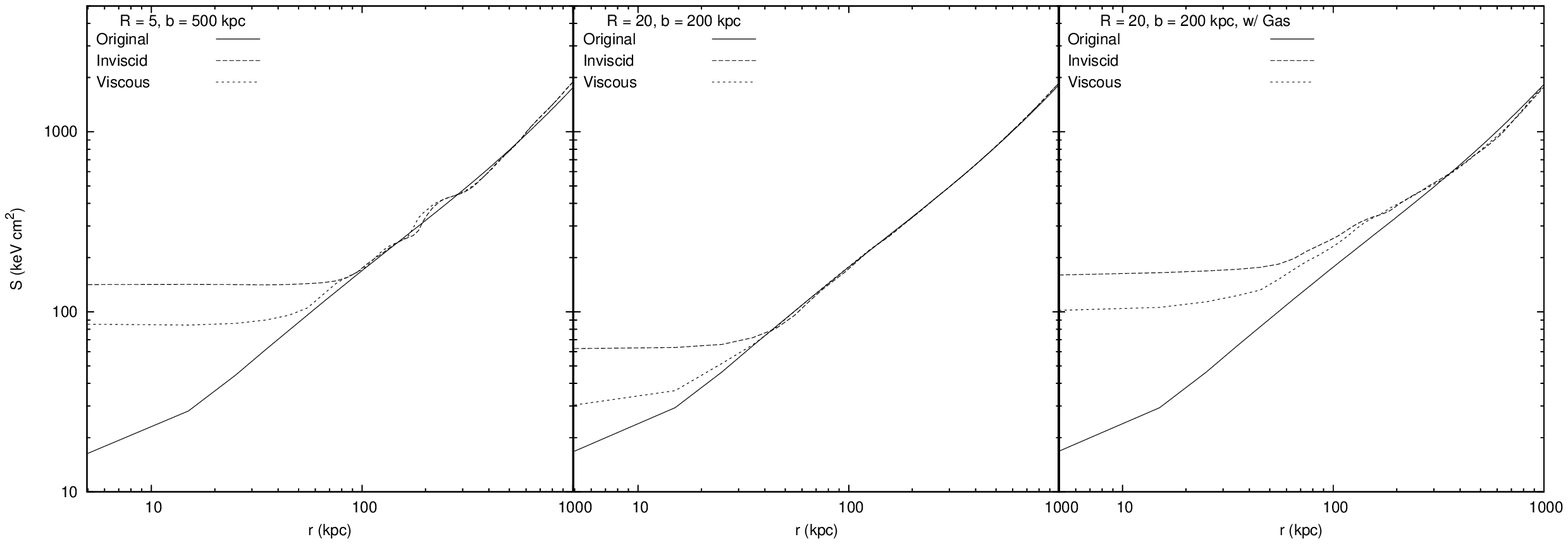}
\caption{Final radial entropy profiles (at $t$ = 6.0 Gyr) for the inviscid and viscous adiabatic simulations.\label{fig:eprofile_visc}}
\end{center}
\end{figure*}

When an explicit physical viscosity is included, its action is to damp gas motions and dissipate them into heat. It might be assumed that increasing the viscosity of the gas would result in more heat, increasing the entropy of the gas further. However, the viscosity has a second effect, which actually results in {\it less} heat being transferred to the cool core than in the case where there is no physical viscosity. Viscosity suppresses instabilities (e.g. Kelvin-Helmholtz) and turbulence, which are the mechanisms that allow for greater mixing of gases with different entropies, and hence greater heating of low-entropy gas (such mixing is the only mechanism for transferring heat from the hot gas to the cool core in our present simulations). Therefore, in the cases where we have included viscosity, the entropy profiles that result have lower entropy floors in the core than the corresponding simulations without an explicit viscosity term. This is shown for simulations R5b500-R20b200gv in Figure \ref{fig:eprofile_visc}, where the profiles of the inviscid and viscous simulations are compared side by side. In each of the relevant cases, the entropy floors of the simulations with viscosity are a factor of $\sim$1.5-2 times lower than those of the inviscid simulations. This result is consonant with the results of \citet{mit09}, who compared binary cluster mergers performed with an Eulerian grid-based code (FLASH) with a Lagrangian smoothed particle hydrodynamics (SPH) code (Gadget). They found that the core entropy floors produced in the FLASH simulations were a factor of $\sim$2 larger than than in the Gadget simulations, and after considering several factors, determined that the larger effective viscosity in the SPH code was responsible for the lower entropy floor, due to its suppression of mixing. This effect is also seen by comparing the stability of the cold fronts in our inviscid, grid-based simulations to the inviscid, SPH simulations of AM06, which had a similar linear resolution in the cluster cores. In that set of simulations, the cold fronts that were produced by sloshing were long-lasting and largely unaffected by K-H instabilities. In our simulations the fronts are quickly disrupted by K-H instabilities, which contribute to the mixing of gases with different entropies that heats up the core gas. We will compare the stability and the appearance of the cold fronts in our simulations and in real clusters in the context of the constraints of plasma viscosity in a separate paper. In the case of simulations R5b500c and R5b500vc, the effect of viscosity made a big difference in terms of how long sloshing would be able to stave off a cooling catastrophe (see Figure \ref{fig:cooling_time}). Under otherwise equal conditions, the action of viscosity in simulation R5b500vc inhibited the mixing of gases of different entropies, and as a result, a cooling catastrophe occurred much earlier ($\Delta{t} \sim$2~Gyr) in simulation R5b500vc than in simulation R5b500c. 

Of course, the viscosity of the collisionless cluster plasma is unknown, and the true value is probably between the cases that we have considered here. Exploring the parameter space of possible values and forms for the viscosity in the ICM will be the subject of a future paper.

\subsection{Compression of the Gas Prior to Sloshing\label{sec:compression}}

\begin{figure*}
\begin{center}
\plotone{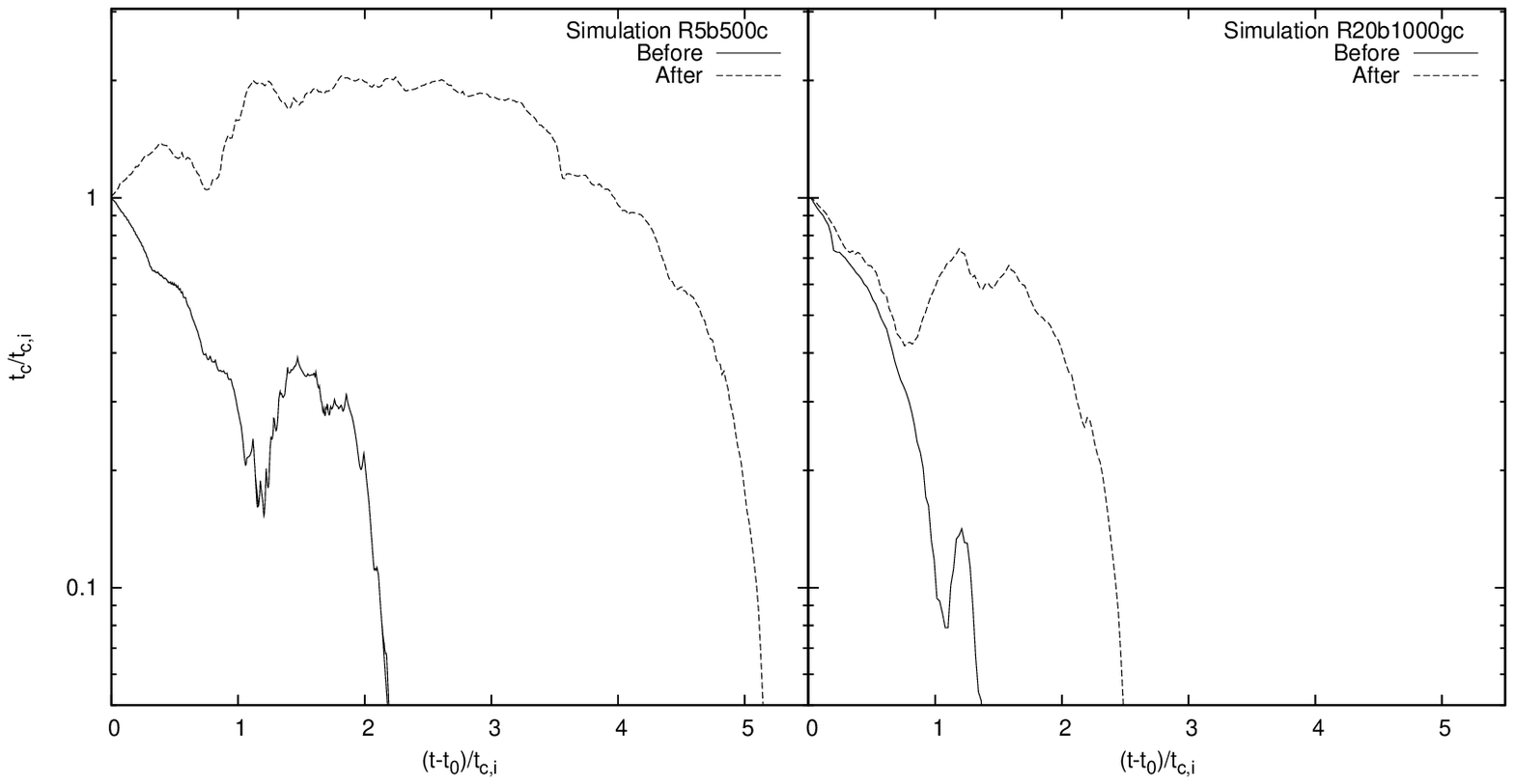}
\caption{The evolution of the minimum cooling time vs. simulation time for simulations R5b500c and R20b1000gc, where cooling is initiated before and after the transient luminosity increase. The simulation time is scaled to the epoch when cooling begins and the cooling time is scaled to units of the initial cooling time.\label{fig:cooling_difference}}
\end{center}
\end{figure*}

One difficulty exists with relying on merging to produce sloshing in cluster cores that may hamper its effectiveness in shutting off a cooling flow. As the subcluster makes its initial passage by the main cluster core, the effect of the increased gravitational potential is to compress the central gas of the main cluster and drive up the luminosity of the core. Additional compression of the core gas occurs if a shock is present (in the simulations with gas in the merging subcluster). Figure \ref{fig:L_vs_time} shows this effect as it occurs in the adiabatic simulations. This increase in luminosity is anywhere from $\sim$1.2-2 times the initial value in the cases where the main cluster gas is merely disturbed, and nearly a fourfold increase in the cases where there is a strong interaction with a gaseous subcluster. This brief increase in luminosity in some cases may be strong enough to initiate a stronger cooling flow that will be more difficult to quench by sloshing. 

The extent to which sloshing is effective in quenching the cooling flow despite the effect of the increase in luminosity can be shown by varying the point at which the cooling is ``switched on'' in a simulation. Figure \ref{fig:cooling_difference} shows the evolution of the central cooling time in simulations R5b500c and R20b1000gc for two different intervals over which cooling is active in the simulation, in one case before the luminosity increase due to compression and in the other afterward. In both cases, there is a significant reduction in the length of time before a cooling catastrophe occurs. In simulation R5b500c, the cooling catastrophe would occur 3 cooling times sooner, and in simulation R20b1000gc, it occurs even earlier than it would have had the cluster not undergone an interaction with a subcluster at all. 

We note here that our initial gas density and temperature profiles describe accurately A2029, which corresponds to a phase where sloshing is active, that is, past the initial luminosity increase, if it had experienced one. We also point out that sloshing may have other causes, which do not have the associated initial increase in luminosity as in the case of a subcluster merger. It is also possible that if the initial cool core was not an ideal, symmetric density peak as assumed in our simulations, but was already fragmented, e.g. by past AGN activity, the increase of the luminosity would be less dramatic.

\section{Summary\label{sec:summary}}

We have performed a set of $N$-body/hydrodynamical simulations of galaxy cluster mergers using the FLASH code to determine whether or not the sloshing of the central cluster gas, caused by such mergers, may provide a source of heat to support the cluster core against developing a strong cooling flow. Our model consists of a large, $T \sim 10$~keV, cool-core cluster (modeled after clusters such as A2029) and a merging subcluster, together in isolation. We have explored a parameter space of possible subcluster encounters, varying the initial mass ratio of the clusters, the impact parameter of the trajectory, and whether or not the subcluster contains gas. Consideration of mergers with gasless subclusters is motivated by the finding by AM06 that such mergers reproduce the frequently observed sloshing in the cores of otherwise relaxed-looking clusters. We have also investigated the effects of varying the ICM viscosity, choosing a simple model where the kinematic viscosity is a constant value approximately equal to the Spitzer value $\sim$50~kpc from the cluster center, and also cases of zero viscosity.

In each simulation, the subcluster approaches the main cluster core, driving a wake (or a shock, if the subcluster contains gas) into the ICM of the main cluster. Though the details of the sloshing process are different in each simulation, the basic cause is the same; as the core of the main cluster is accelerated by the gravitational tug of the subcluster, the ram pressure of the surrounding intracluster medium pushes the core gas out of equilibrium with the dark matter core, and after the effect of this ram pressure subsides, this gas falls back into the dark matter core and begins to ``slosh'' back and forth in the gravitational potential well. If the incoming subcluster has a non-zero impact parameter, the wake (or shock) produced by the subcluster transfers angular momentum to the core gas, resulting in a spiral-shaped pattern of sloshing. 

One set of simulations was performed without radiative cooling, in order isolate the effects of sloshing. In each case, hot gas from larger radii in the cluster was mixed with the cool gas of the core, resulting in a net increase of the temperature and entropy of the core gas. A relevant side effect of this heating is that the core gas also expands, decreasing the rate of radiative cooling (due to its strong dependence on the density). In all but the weakest-disturbance cases, the effect of this sloshing is to increase the central entropy of the gas core by a factor of nearly $\sim$3-10. It was also seen that the effect of including viscosity in the simulations is to damp out fluid instabilities and turbulence, which results in far less mixing of the core gas. As a result, the amount of entropy increase of the core is reduced. However, if the infalling subcluster contains gas, the amount of heat input to the cluster core is significantly increased, due to the associated shock heating and mixing induced by a much stronger hydrodynamic disturbance.

We performed a second set of simulations, restarting four of our most interesting adiabatic cases and switching on the effects of radiative cooling. When cooling is included in the simulations, sloshing still occurs, but now must compete against the radiative losses and the resulting gas density increase to maintain the density and temperature structure of the cluster core. We find that, as should be expected, the merger setups that resulted in the strongest sloshing in the corresponding adiabatic simulations manage to offset the effects of cooling for the longest intervals of time. This is seen clearly in the increase of the central cooling time by nearly a factor of $\sim$2-4. Sloshing suppresses the increase of the mass of gas with low cooling times, in stark contrast to the case where a cluster remains undisturbed. In each case, however, the sloshing in our idealized simulations eventually becomes weak and a strong cooling flow develops. Depending on the strength of the sloshing, we demonstrated that a cooling catastrophe can be prevented for intervals of time $\sim$1-3~Gyr. If encounters with subclusters are frequent enough, subsequent merging activity may be able to pick up where the last subcluster ``left off'' to continue to drive the sloshing of the cluster core. 

Though we have considered the process of sloshing in isolation, in reality the cores of galaxy clusters host a variety of other processes, such as energy injection from active galactic nuclei and supernovae. These effects would also act to heat the cluster core. A full treatment of the strength of the various heating mechanisms present in cluster cores and the sum of these effects will require further simulations. Future work will also have to focus on modelling more accurately the viscosity of the ICM, which is likely not only to have a dependence on the density and temperature of the gas but also is inevitably sensitive to the properties of the cluster magnetic field. In addition, thermal conduction due to electron collisions, much too weak to stave off cooling by itself, may help significantly in the presence of sloshing, which brings hot and cool gas phases in close contact. This will be explored in a future paper.

\acknowledgments
Calculations were performed using the computational resources of Lawrence Livermore National Laboratory, Argonne National Laboratory, the Texas Advanced Computing Center, and the National Institute for Computational Sciences. Analysis of the simulation data was carried out using Nathan Hearn's QuickFlash tools, which are available for download at \url{http://quickflash.sourceforge.net}. JAZ is grateful to Paul Ricker and Paul Nulsen for useful discussions and advice, and to Yago Ascasibar for providing the initial conditions from AM06. JAZ is supported under {\it Chandra} grant GO8-9128X. MM was supported by NASA contract NAS8-39073 and the Smithsonian Institution. The software used in this work was in part developed by the DOE-supported ASC / Alliances Center for Astrophysical Thermonuclear Flashes at the University of Chicago.

\appendix

\section{Resolution Test}

Crucial to our hypothesis that heating from sloshing may be able to offset cooling is that the hot gas from higher radii in the cluster will mix with the cold gas from the core region, increasing the entropy of the core. Mixing is aided by the growth of instabilities, particularly the Kelvin-Helmholtz (K-H) instability. As the simulations demonstrated, viscosity will damp these instabilities, inhibiting mixing and resulting in less heating of the cluster core. Even without an explicit viscosity, however, the finite resolution of the simulations imposes a lower limit on the effective numerical viscosity of the gas. The dynamic coefficient of viscosity corresponding to the resolution elements is of order $v\Delta{x}$, where $\Delta{x}$ is the size of the resolution element and $v$ is the characteristic speed of the gas motion. Put another way, the finite resolution of the simulations places a floor on the wavelengths of the unstable perturbations that can be resolved on the grid. Consequently, it may be of some concern that we are underestimating the amount of heat generated via mixing by our inability to resolve the full range of wavelengths of perturbations that can become unstable. Additionally, since the entropy declines with radius as a power-law essentially all the way to the cluster core, the inability to resolve radii less than the resolution size places an effective floor on the initial entropy profile. This effect would overestimate the heat generated due to an initial overestimation of the core entropy. 

It is prohibitive to test for convergence of our results within our standard simulation setup by running simulations of progressively higher resolution due to the high expense of the gravitational potential calculation in the current realization of FLASH. Therefore, to check against resolution effects, we have ran tests using a simplified model for our binary merger setup. The underlying physical model is the same, with the exception that the gravitational potential for the main cluster and subcluster are modeled as rigid fields (i.e., the cluster DM distributions are considered to be solid collisionless bodies). The potential field of the subcluster begins at the same position as in the self-gravitating case, and has a trajectory that is calculated assuming it falls as a point mass within the potential of the main cluster. The main cluster's potential is held fixed at the center of the grid domain, but since this is not an inertial frame, we calculate the corresponding inertial acceleration felt by the gas in this frame and add it to the gravitational acceleration from the subcluster. The sloshing created by this encounter is slightly different in detail, but qualitatively the same to that of the self-gravitating model; in any case, our concern is to demonstrate the convergence of our result with respect to the entropy increase of the core with increasing resolution, which will only depend on the details of the hydrodynamics. This setup will be used to analyze aspects of the sloshing process in more detail in a forthcoming paper.

\begin{figure}
\begin{center}
\plotone{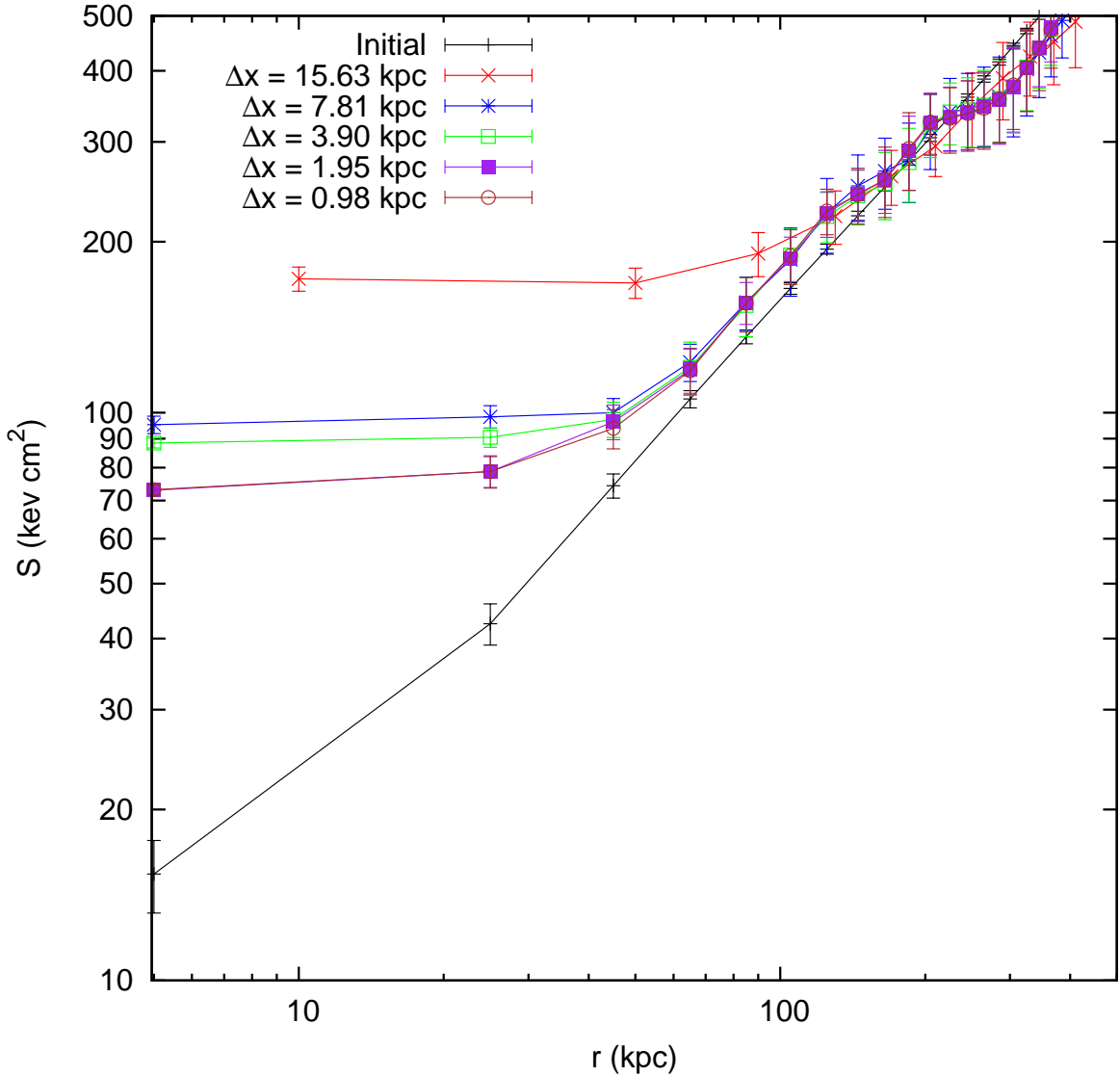}
\caption{Plot of gas entropy vs. radius for a series of rigid-potential simulations with differing resolution.\label{fig:res_test}}
\end{center}
\end{figure}

For the purpose of a resolution convergence test, a toy model of simulation R5b500 was ran with four different levels of resolution. Our lowest resolution is $\sim$16~kpc, corresponding to $\sim3$x the resolution of our self-gravitating simulations, and our highest resolution is $\sim$1~kpc. The central entropy profile is raised and flattened as in the self-gravitating simulations. Therefore we are able to resolve perturbations with wavelengths that are 5 times smaller than in the self-gravitating simulations. Figure \ref{fig:res_test} shows the resulting entropy profiles of the different resolution simulations, compared to the initial entropy profile. For cell sizes $\Delta{x} \simlt$ 8~kpc, the entropy profile is fairly well-converged, whereas for larger cell sizes the entropy floor of the profile is higher as the initial core entropy is overestimated due to the low resolution. We conclude that our results from the self-gravitating simulations presented in this work have not been significantly affected by resolution effects.

\section{Modifying the Central Potential}

Most clusters of galaxies of the type we have been concerned with in this study, namely those of the ``cool-core'' variety, harbor a central massive cD galaxy (Jones \& Forman 1984; Peres et al.\ 1998). The gravitating mass from the dark and baryonic components of this galaxy will alter the shape of the potential in the central regions. The important effect for our purposes will be to deepen the potential well, making it more difficult to displace the cold central gas from the center and mix it in with hot gas from the outskirts. To determine if this would have a significant effect on our conclusions, we have set up a rigid-potential simulation (the same setup as in Appendix A for our resolution test) and added a potential component representing a cD galaxy to the potential component representing the main cluster's dark matter, placing it at the center of the latter's potential well. Recent modeling of observed cD galaxy gravitational potentials \citep[e.g.][]{chu10,hum10} have shown that they are well-represented by an isothermal sphere potential (with $\rho(r) \propto r^{-2}$ and $\Phi(r) \propto \log{r}$). To approximate this potential but avoid the unphysical behavior of this form at the center, we adopt the softened isothermal sphere model, $\rho(r) \propto [1+(r/r_c)^2]^{-1}$, with a small core radius $r_c$ of 0.1~kpc. We chose a mass $M_{\rm cD} = 5.0 \times 10^{12} M_{\odot}$ and a radius $R_{\rm cD} = 100~{\rm kpc}$ for the galaxy. 

\begin{figure}
\begin{center}
\plotone{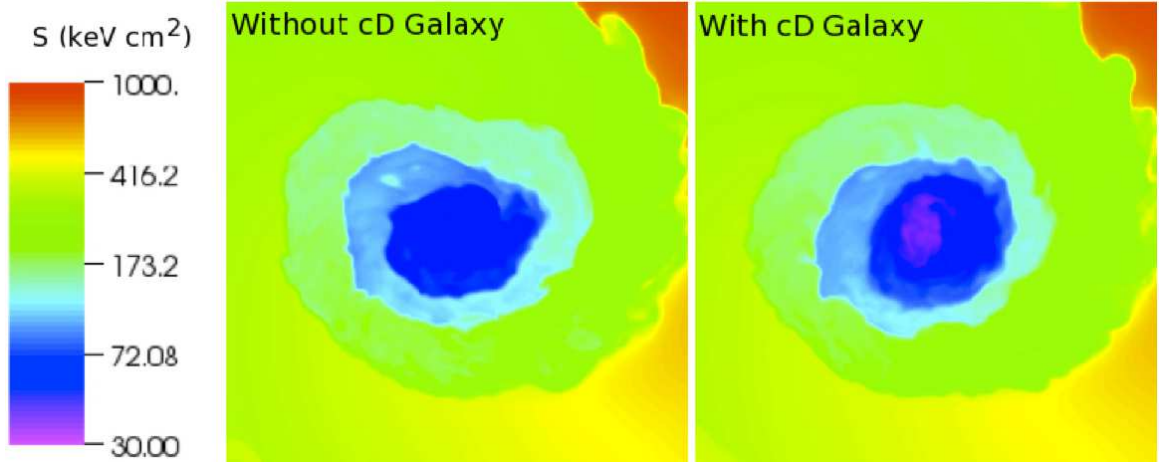}
\caption{Slices of the entropy through the center of the domain for cases with and without a central cD galaxy, shown at two different times in the simulation.\label{fig:cD_entr_fig}}
\end{center}
\end{figure}

\begin{figure}
\begin{center}
\plottwo{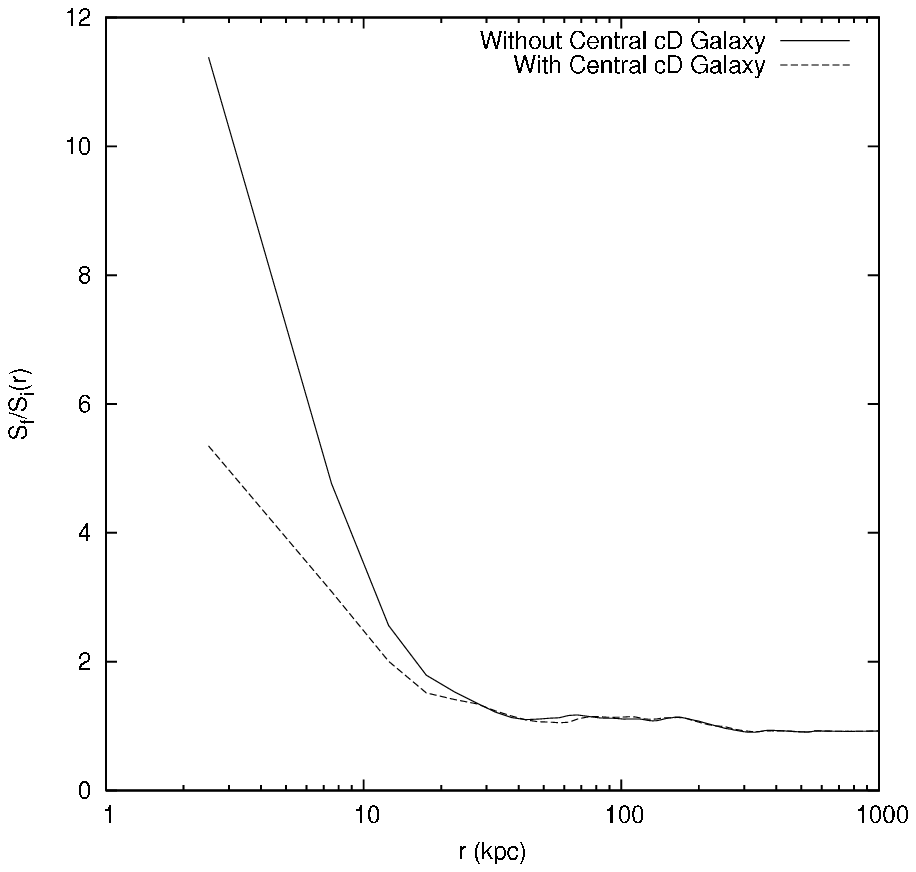}{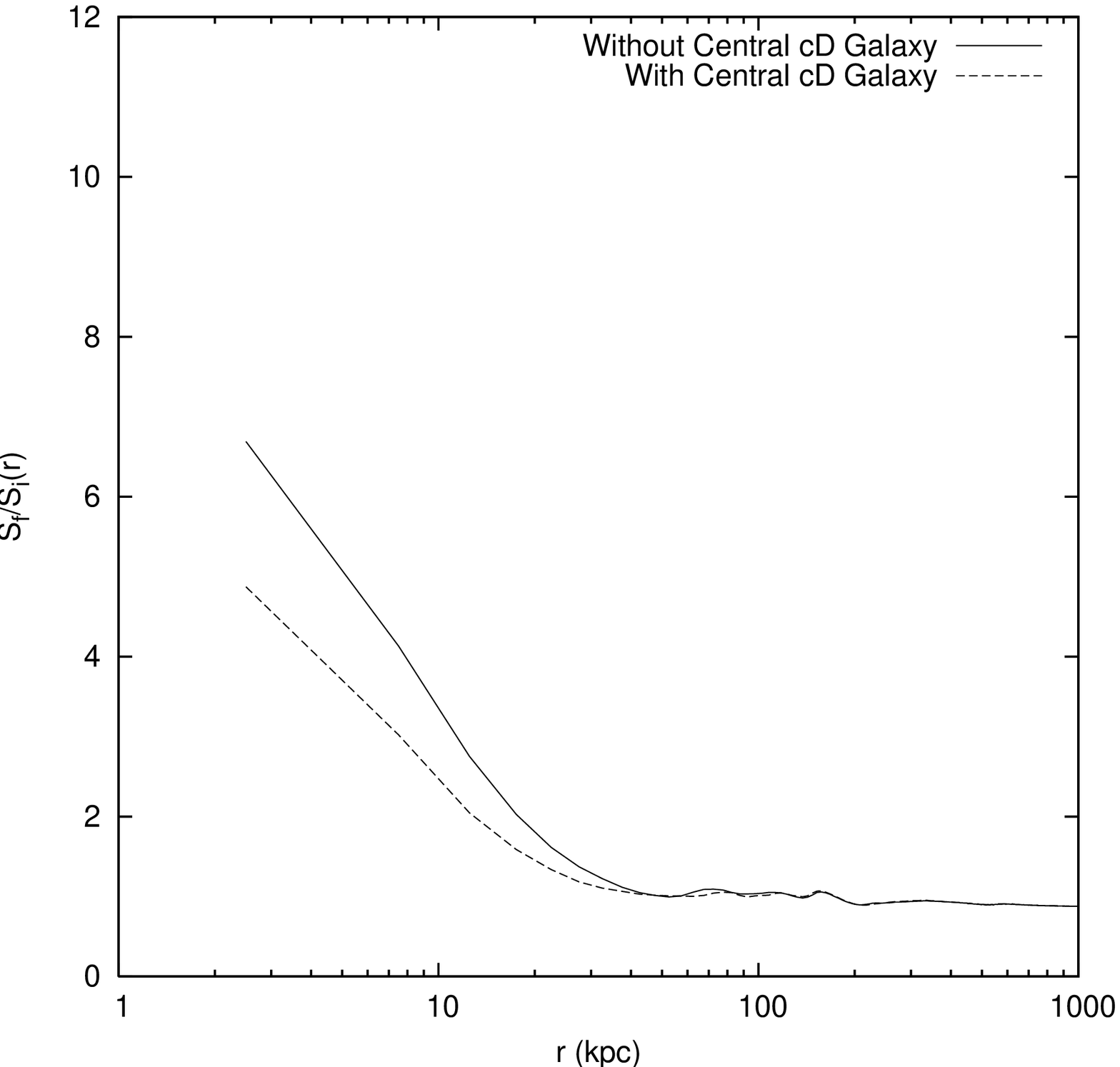}
\caption{Ratio of the intial and final entropy radial profiles vs. radius for sloshing simulations with and without an additional potential component due to a cD galaxy. Left: The case where the initial gas profile is the same in both potential cases. Right: The case where the initial entropy profile is the same in both potential cases.\label{fig:cD_entr_ratio}}
\end{center}
\end{figure}

Figure \ref{fig:cD_entr_fig} shows the effect of steepening the central potential on the entropy distribution of the main cluster for a merger simulation with $R$ = 5, $b$ = 500~kpc. A case without a cD galaxy potential component is compared to a case with a cD galaxy potential, with the initial entropy profile constructed to be the same for both cases. The presence of the cD galaxy results in it being more difficult to push the low-entropy gas out of the potential, resulting in the gas maintaining a lower entropy at the center after sloshing has disrupted the core. Figure \ref{fig:cD_entr_ratio} further demonstrates this effect by comparing the ratio of the final to the initial radial entropy profiles in the case where there is a cD galaxy and the case where there is no cD galaxy at the center. We have examined this effect in two cases, one where the gas density profile is the same in both potential cases and another where the gas entropy profile is the same. In all cases, the core entropy is increased by a factor of at least several. However, in the presence of the cD galaxy the increase of core entropy is $\sim 1.4-2\times$ smaller when compared to the case where there is no central galaxy. This indicates that in realistic clusters with cD galaxies, interactions with subclusters will still be able to perturb the cluster core enough to cause sloshing and mixing with higher-entropy gas, though the effectiveness of this mechanism to heat the core will be diminished. The difference between the resulting average central entropies in our test run with and without a cD galaxy is comparable to the differences between our other simulations with different merger parameters. Thus, sloshing may still be an effective mechanism for heating the core if it is strong enough.

\end{document}